\documentclass[12pt]{iopart}
\usepackage[utf8]{inputenc}

\expandafter\let\csname equation*\endcsname\relax
\expandafter\let\csname endequation*\endcsname\relax

\usepackage{amsmath}
\usepackage{graphicx}
\usepackage{float}
\usepackage{wrapfig}
\usepackage{braket}
\usepackage{bm}
\usepackage{cleveref}
\usepackage[most]{tcolorbox}
\usepackage[
backend=biber,
style=phys,
sorting=none
]{biblatex}
\tcbset{textmarker/.style={%
        enhanced,
        parbox=false,boxrule=0mm,boxsep=0mm,arc=0mm,
        outer arc=0mm,left=6mm,right=3mm,top=7pt,bottom=7pt,
        toptitle=1mm,bottomtitle=1mm,oversize}}

\newtcolorbox{noteBox}{textmarker,
    borderline west={6pt}{0pt}{green},
    colback=green!10!white}

\newcommand{\q}{\mathbf{q}}
\newcommand{\kk}{\mathbf{k}}
\newcommand{\rr}{\mathbf{r}}

\NewDocumentCommand{\code}{v}{%
\texttt{\textcolor{black}{#1}}%
}
\usepackage{multirow}

\begin{document}
\title[Ab initio electronic friction]{Ab initio calculation of electron-phonon linewidths and molecular dynamics with electronic friction at metal surfaces with numeric atom-centered orbitals}
\author{Connor L. Box, Wojciech G. Stark, Reinhard J. Maurer}
\address{Department of Chemistry, University of Warwick, Gibbet Hill Rd, Coventry, CV4 7AL, UK}
\date{\today}

\begin{abstract}
Molecular motion at metallic surfaces is affected by nonadiabatic effects and electron-phonon coupling. The ensuing energy dissipation and dynamical steering effects are not captured by classical molecular dynamics simulations, but can be described with the molecular dynamics with electronic friction method and linear response calculations based on density functional theory. Herein, we present an implementation of electron-phonon response based on an all-electron numeric atomic orbital description in the electronic structure code FHI-aims. After providing details of the underlying approximations and numerical considerations, we present significant scalability and performance improvements of the new code compared to a previous implementation [Phys. Rev. B 94, 115432 (2016)]. We compare convergence behaviour and results of our simulations for exemplary systems such as CO on Cu(100), H$_2$ adsorption on Cu(111), and CO on Ru(0001) against existing plane wave implementations. We find that our all-electron calculations exhibit faster and more monotonic convergence behaviour than conventional plane-wave-based electron-phonon calculations. Our findings suggest that many electron-phonon linewidth calculations in literature may be underconverged. Finally, we showcase the capabilities of the new code by studying the contribution of interband and intraband excitations to vibrational linewidth broadening of aperiodic motion in previously unfeasibly large periodic surface models.
\end{abstract}

\maketitle

%\tableofcontents

%%%%%%%%%%%%%%%%%%%%%%%% here it is %%%%%%%%%%%%%%%%%%%%%%%%%%%%%%%%%
%%%%%%%%%%%%%%%%%%%%%%%% here it is %%%%%%%%%%%%%%%%%%%%%%%%%%%%%%%%%

% ================================================================ %
                    \section{Introduction}
% ================================================================ %

%PAR 1
Electronic and vibrational dynamics in metallic systems are strongly coupled, ~\cite{Hellsing2002, kroger_electronphonon_2006} due to the insufficient energy scale separation of low-lying electronic excitations and vibrational motion.  This also holds true for the reactive dynamics of molecules adsorbed at metallic surfaces.~\cite{auerbach_chemical_dynamics_2021} Strong electron-phonon coupling leads to nonadiabatic effects on adsorbate dynamics at metal surfaces with experimentally measurable consequences. This includes the ultrafast decay of adsorbate vibrational excitation, measurable as picosecond-scale vibrational lifetimes of molecular adsorbates.~\cite{Persson1982, Arnolds2011} Ultrashort adsorbate lifetimes have been reported for systems such as carbon monoxide (CO) on Cu(100),~\cite{Morin1992, Hirschmugl1990, Hirschmugl1994,Inoue2016} and on Pt(111).\cite{Beckerle1991} Molecular beam scattering experiments of hydrogen atoms and diatomics such as nitric oxide have reported highly vibrationally inelastic scattering behaviour that arises from the excitation of electrons due to atomic and molecular impingement \cite{Bunermann2015, huangVibrationalPromotionElectron2000, Rahinov2011}. Equally, ultrafast time-resolved spectroscopy measurements have shown efficient activation of molecular motion and desorption driven by hot electron excitation.\cite{Inoue2016, diesen_ultrafast_2021} Beyond capturing the fundamental dynamics at surfaces, hot electron effects are also discussed in the context of light-controlled chemistry and catalysis.\cite{Park2015, zhangSurfacePlasmonDrivenHotElectron2018}

%PAR 2
The accurate and efficient simulation of electron-phonon coupling (EPC)  is crucial to provide insights on vibrational lifetimes and linewidths, \cite{Bauer1998} critical superconducting temperatures and electron-phonon instabilities in phonon spectra. \cite{giustinoElectronphononInteractionsFirst2017} Several software packages have been established for the calculation of EPC effects in condensed matter based on pseudopotential plane-wave formalisms (e.g. {\sc Quantum ESPRESSO} \cite{giannozziAdvancedCapabilitiesMaterials2017}, EPW \cite{noffsingerEPWProgramCalculating2010} or YAMBO \cite{Marini2009}). Most existing approaches perform perturbative expansions starting from the harmonic phonon spectrum, ie. they calculate the frequency renormalization or linewidth associated with EPC. \cite{giustinoElectronphononInteractionsFirst2017,Novko2016} In order to deliver converged and robust results, very dense Brillouin zone integration for both electrons (``$\kk$-space") and phonons (``$\q$-space") is required, which is often achieved by transformation to a maximally localized Wannier basis for interpolation of coupling matrix elements. \cite{giustinoElectronphononInteractionUsing2007, agapitoInitioElectronphononInteractions2018} Most phonon linewidth calculations in literature have been calculated following this approach based on the simplified first-order perturbation theory expression introduced by Allen and coworkers (\textit{vide infra}). \cite{Allen1972}

%PAR 3
The simulation of EPC and nonadiabatic energy dissipation effects in chemical dynamics at molecule-metal interfaces requires molecular surface models that correctly capture physical coverage and bonding scenarios, and that explore dynamics well beyond the harmonic regime. In the case of low coverage adlayers or catalytic models, this can lead to systems featuring hundreds of atoms for which the dynamical phase space needs to be explored. This provides a significant computational challenge for existing plane-wave implementations. Only recently have atomic-orbital based EPC implementations emerged. \cite{frederiksenInelasticTransportTheory2007, zhouPerturboSoftwarePackage2021} Atomic orbital based codes offer the benefit of a sparse and atom-centred basis, which leads to compact matrix representations of the Hamiltonian. The FHI-aims code is a rare example of an all-electron full-potential-based numeric atom-centered orbital (NAO) implementation of electronic structure theory based on density functional theory.\cite{Blum2009} Recently within FHI-aims, Shang et al. have developed a linear response implementation for the calculation of phonons and electric field response. \cite{shangLatticeDynamicsCalculations2017} FHI-aims is highly scalable on high performance architectures due to the use of distributed matrix parallelism as provided by the eigenvalue solvers for petaflop applications (ELPA) library \cite{Marek2014} that is part of the electronic structure infrastructure (ELSI) project. \cite{yuELSIUnifiedSoftware2018}

\begin{figure}
    \centering
    \includegraphics[width=6in]{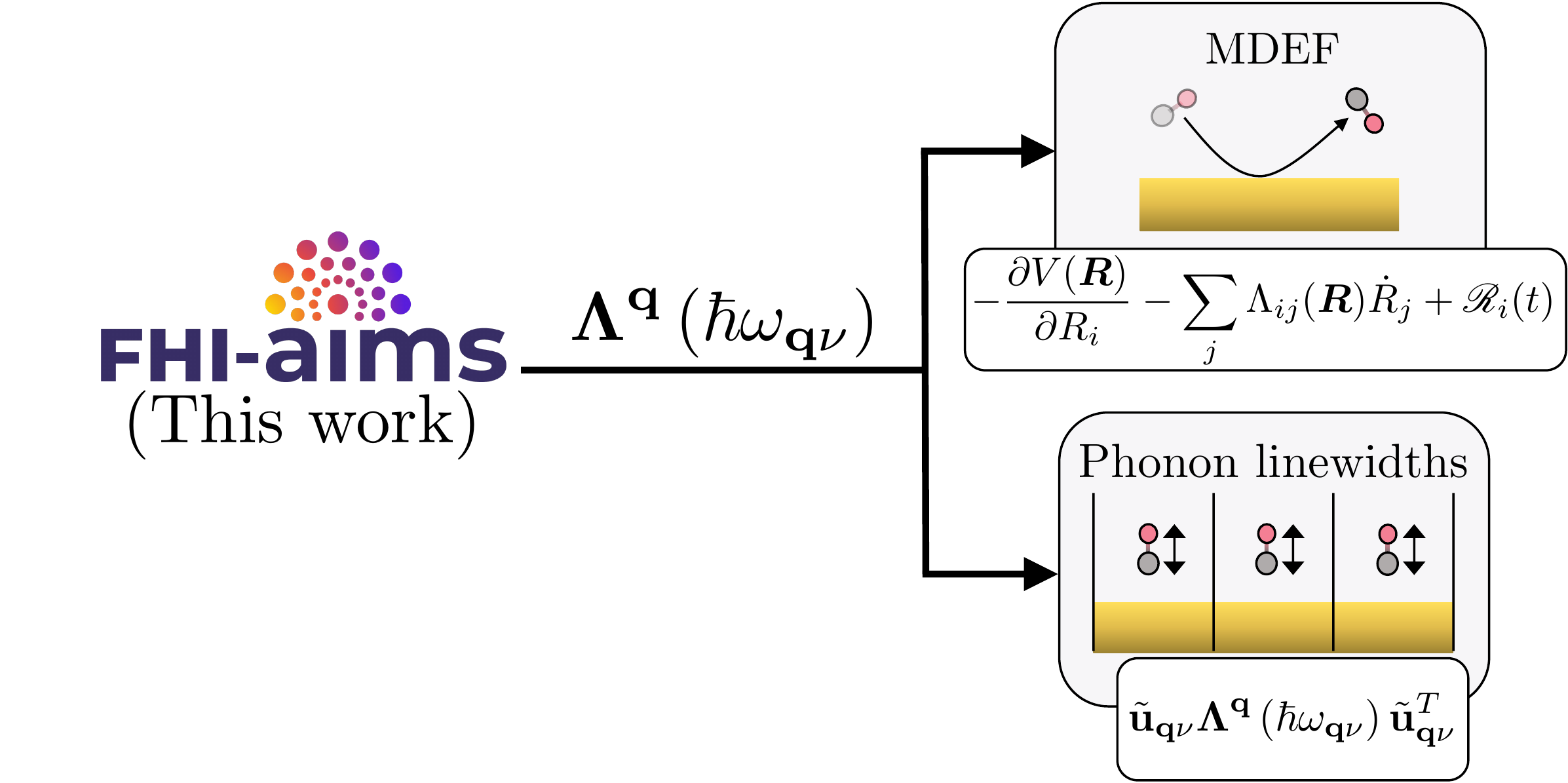}
    \caption{Conceptual graphic. FHI-aims software can calculate the position ($\bm{R}$) and energy ($\hbar\omega$) dependent electronic friction tensor, $\Lambda^{\mathbf{q}}\left(\hbar \omega_{\mathbf{q} \nu}\right)$ using DFT. The tensor can be used in MDEF (we show a molecule scattering from a metal surface) or projected along normal mode displacements $\mathbf{u}_{\q\nu}$ to calculate phonon linewidths. We mostly focus on the latter in this work, but the implementation serves both applications.}
    \label{fig:workflow}
\end{figure}

%PAR 4
EPC effects on dynamics can also be represented in an anharmonic or Cartesian representation as an alternative to describing EPC effects in the harmonic regime based on phonons. This is much less common but has the benefit of enabling an anharmonic treatment of transport and dissipation effects. \cite{Maurer2017,luSemiclassicalGeneralizedLangevin2019, zachariasFullyAnharmonicNonperturbative2020} One such approach is the molecular dynamics with electronic friction (MDEF) method, where the dynamics of nuclei are represented via classical equations of motion (see MDEF in \cref{fig:workflow}, top). \cite{Head-Gordon1995, Persson1982, douBornOppenheimerDynamicsElectronic2017} In MDEF, forces on the nuclei derive from the potential energy landscape that arises from the ground-state electronic structure and additional force contributions that relate to dissipative interactions due to hot electron excitations in the metal. The central quantity that governs these forces is the friction tensor $\mathbf{\Lambda}$, which is a function of atomic coordinates and can be directly calculated from EPC based on first-order perturbation theory.\cite{Askerka2016} Maurer et al. have presented an implementation of electronic friction based on linear response calculation of EPC in FHI-aims.\cite{Maurer2016a} This implementation has enabled the calculation of on-the-fly MDEF simulations based on DFT for reactive scattering of H$_2$ on Ag(111) and has subsequently been used to construct machine-learning-based representations of the electronic friction tensor for state-to-state scattering of H$_2$ on Ag(111) \cite{Maurer2017, zhangHotelectronEffectsReactive2019,maurerHotElectronEffects2019} and NO on Au(111).\cite{boxDeterminingEffectHot2020} The electronic friction tensor can also be employed to directly calculate EPC-based vibrational linewidths equivalent to the conventional approach, by simply projecting the Cartesian friction tensor into normal mode space as shown in  \cref{fig:workflow}, where $\tilde{\mathbf{u}}_{\q\nu}$ refer to mass-weighted phonon displacement eigenmodes associated with phonon frequency $\omega_{\q\nu}$. \cite{Maurer2016a}

%PAR 5
In this work, we present a detailed introduction and assessment of ab initio electronic friction calculations based on an all-electron NAO implementation of DFT. We present a refactorization of the linear response EPC and electronic friction code in FHI-aims that delivers a substantial increase in computational efficiency and scalability. This code enables the calculation of EPCs, electronic friction tensors, and EPC-based vibrational linewidths. Details of the underlying theory, the necessary approximations and numerical choices are provided. We show that the code delivers reliable and converged predictions with improved convergence behaviour compared to existing plane wave pseudopotential EPC implementations. The improved scalability of the code enables the study of periodic systems with more than 250 atoms per unit cell, which we showcase by calculating the inter- and intraband contributions to the vibrational linewidth of aperiodic internal stretch vibration in a coherent c($2\times2$) CO monolayer adsorbed at Cu(100).

% ================================================================ %
                        \section{\label{sec:theory}Theory}
% ================================================================ %

\subsection{Nonadiabatic effects on lattice vibrations in metals}
% ---------------------------------------------------------------- %

Many-body perturbation theory provides a general framework to understand how EPC and nonadiabaticity affect phonon properties.~\cite{Mahan2013,giustinoElectronphononInteractionsFirst2017} The harmonic phonon spectrum is  based on noninteracting phonon frequencies $\omega_{\q\nu}$ that are calculated from DFT via the dynamical matrix of the system. \cite{Baroni2001} This approach already considers some effects of electron-electron correlation on the frequency of the phonon. The phonon is characterized by two quantum numbers, namely $\nu$, which refers to the band index of the phonon and $\q$, which refers to the crystal periodic momentum vector. Phonon frequencies correspond to the poles on the frequency axis of the adiabatic one particle phonon Green's function, $D^A_{\nu}(\q)$\cite{Novko2016, giustinoElectronphononInteractionsFirst2017}:
\begin{equation}\label{eq:phonon-self-energy-adiabatic}
    D_{\nu}^A(\q,i\omega_n) = \frac{2\omega_{\q\nu}}{(i\omega_n)^2-\omega_{\q\nu}} 
\end{equation}
Where $i\omega_n$ is a complex variable to be replaced with $\omega+i\eta$ when retarded expressions are formed for practical evaluations. $\eta$ is a positive real infinitesimal regularization constant. The phonon spectrum is given by the spectral function
\begin{equation}
    B^A(\q,\omega) = -2\sum_{\nu}\mathrm{Im}\left[  D_{\nu}^A(\q,i\omega_n) \right] = 2\pi \sum_{\nu}\delta(\omega - \omega_{\q\nu})
\end{equation}
Each frequency corresponds to a delta resonance of infinite lifetime.

%general phonon greens function Dyson equation
The interacting phonon Green's function $D(\q,\omega)$ can be written as\cite{Novko2018}
\begin{equation}
    D(\q,i\omega_n) = \sum_\nu \frac{2\omega_{\q\nu}} {(i\omega_n)^2+\omega_{\q\nu}^2+2\omega_{\q\nu} \Pi^{\mathrm{NA}}_{\nu}(\q,i\omega_n)}
\end{equation}
The effect of EPC is described by the nonadiabatic electron-phonon self-energy $\Pi_{\nu}^{\mathrm{NA}}(\q,i\omega_n)$. EPC can lead to a shift (renormalization) of phonon frequencies, coupling of phonons, and to linewidth broadening by introducing a finite lifetime to the phonon. The real part of $\Pi^{\mathrm{NA}}_{\nu}(\q,i\omega_n)$ is associated with frequency renormalization and the imaginary part with linewidth broadening and lifetime of the phonon. The interacting Green's function $\mathbf{D}(\omega)$ has complex poles $\Tilde{\Omega}_{\q\nu}$ of $D^{-1}_{\q\nu\nu}(\omega)$ (i.e. frequencies at which $D^{-1}=0$) which have a real and a complex contribution $\Tilde{\Omega}_{\q\nu}^2=\Omega_{\q\nu}^2 - i\gamma_{\q\nu}$. The effects of $\Pi_{\nu}^{\mathrm{NA}}(\q,i\omega_n)$ on the phonon spectrum are complex and involve mode coupling and multiple poles for the same frequency, leading to new structures in the phonon spectrum. Almost all practical EPC calculations only consider diagonal contributions to the linewidth, i.e. they ignore mode coupling and multiple pole structures. The spectral function of the nonadiabatic interacting phonon spectrum is typically approximated as a sum of quasiparticle peaks represented by a Lorentzian centred around the renormalized frequency $\Omega_{\q\nu}^2$ with a full width at half maximum (FWHM)  of $\gamma_{\q\nu}$: \cite{giustinoElectronphononInteractionsFirst2017}
\begin{equation}\label{eq:ep-broadening}
    \gamma_{\q\nu} = -2\frac{\omega_{\q\nu}}{\Omega_{\q\nu}} \mathrm{Im}\Pi^{NA}_{\q\nu\nu}(\Omega_{\q\nu}-i\gamma_{\q\nu}) \approx  -2\mathrm{Im} \Pi^{NA}_{\q\nu\nu}(\omega_{\q\nu})
\end{equation}

%%practical calculations of nonadiabatic corrections 
Many approximations are applied in practical applications. For example the EPC phonon FWHM, $\gamma_{\q\nu}$ is most often evaluated only up to first-order in electron-phonon coupling with first-order perturbation theory and only diagonal corrections are applied: $\q=\q'$ and $\nu=\nu'$. \cite{giustinoElectronphononInteractionsFirst2017} This leads to the neglect of off-diagonal contributions of $\Pi^{NA}_{\q\nu\q'\nu'}(\omega)$, which we will partially introduce below when presenting an expression in Cartesian representation that depends on the Cartesian electronic friction tensor.

EPC-based vibrational linewidth calculations are typically performed in the single particle picture given by independent Kohn-Sham electrons within DFT. Herein, certain approximations are implicit (e.g. ignoring vertex, replacing interacting with noninteracting one-electron Green's function, replacing dielectric matrix with RPA+xc response obtained from DFT, neglecting frequency dependence of coupling elements, $g$ \cite{giustinoElectronphononInteractionsFirst2017}). In this framework, the first-order contribution to the electron-phonon self-energy $\Pi^{NA}_{\nu\nu'}$ can be expressed as:
\begin{equation}\label{eq:NA_selfenergy}
    \hbar\Pi^{NA}_{\nu\nu'}(\q,\omega) = 2\sum_{mn}\int \frac{d\kk}{\Omega_{BZ}} g_{mn\nu}^b(\kk,\q)g_{mn\nu'}^*(\kk,\q) \times \left[ \frac{f_{m\kk+\q}-f_{n\kk}}{\epsilon_{m\kk+\q}-\epsilon_{n\kk}-\hbar(\omega+i\eta)} - \frac{f_{m\kk+\q} - f_{n\kk}}{\epsilon_{m\kk+\q}-\epsilon_{n\kk}} \right]
\end{equation}
In \cref{eq:NA_selfenergy}, $f_{n\kk}$ refers to occupations of electronic states, $g^b$ and $g$ correspond to bare and screened EPC. Due to momentum conservation in periodic systems, contributions with $\q'\neq\q$ vanish at first-order. The factor of 2 arises from spin degeneracy. It is common practice to replace the bare matrix element $g^b_{mn\nu}(\kk,\q)$ with the screened element $g_{mn\nu}(\kk,\q)$, which is the standard output from linear response theory based on DFT.\cite{Allen1972} Calandra, Profeta, and Mauri have argued that this replacement is justified as the error is of second order in the induced electron density, hence it can be neglected.\cite{calandraAdiabaticNonadiabaticPhonon2010} We note here that very recently this standard practice has been assessed by Paleari and Marini who have concluded that this has the potential to be a significant source of error in the calculation of linewidth broadening. \cite{paleariElectronPhononInteraction2021, paleariOverscreeningfreeElectronphononInteraction2021} We take note of this caveat, but will use the standard ``overscreened" approach to EPC in this work as it is the current \textit{status quo} in all existing software packages. A full assessment of this overscreening effect for the calculation of adsorbate vibrational linewidth calculations will be a matter of future investigations.

%broadening, rate, lifetime
Often, experimental values are reported in terms of FWHM broadening, vibrational relaxation rates, or vibrational lifetimes. The vibrational lifetime is related to the FWHM via the Heisenberg uncertainty: $\tau_{\q\nu}=\hbar/\gamma_{\q\nu}$. The relaxation rate in inverse time is defined as $\Gamma_{\q\nu}=\gamma_{\q\nu}/\hbar$.

\subsection{The electronic friction tensor}
% ---------------------------------------------------------------- %

%%%FIRST WE WRITE DOWN EXPLICIT EXPRESSIONS FOR RELAXATION RATE ALONG PHONON MODES

%single delta expression
A common practical expression for the vibrational linewidth of a phonon, $\gamma_{\q\nu}$ as defined in eq.~\ref{eq:ep-broadening} induced by first-order electron-phonon coupling is:
\begin{equation}\label{eq:first_order}
    \gamma_{\q\nu}(\hbar\omega_{\q\nu}) = 2\pi\sum_{\sigma m n}\int \frac{d\kk}{\Omega_{BZ}} |g_{mn\nu}(\kk,\q)|^2 (f_{n\kk}-f_{m\kk+\q}) \delta(\epsilon_{m\kk+\q}-\epsilon_{n\kk}-\hbar\omega_{\q\nu})
\end{equation}
This expression can be obtained from the imaginary part of the first-order nonadiabatic phonon self-energy (\cref{eq:NA_selfenergy}) \cite{giustinoElectronphononInteractionsFirst2017, Novko2018} or can be derived from Fermi's golden rule.\cite{albersNormalUmklappPhonon1976,Maurer2016a} Eq.~\ref{eq:first_order} covers only interband excitations for the case of $\q=0$ ($m>n$) and interband and direct intraband excitations for the case of $\q\neq 0$ ($m\geq n$). Higher order electron-phonon effects (1-electron, 2-phonon scattering) can lead to additional intraband contributions, but they are not covered by this expression.\cite{Novko2018,Novko2016} We will indirectly incorporate such effects in section~\ref{sec:intraband} by band folding in larger unit cells. In this and the following expressions, we explicitly note the sum over spin, $\sigma$.

%from single delta to double delta, zero frequency / low T approximation
The integration of the Dirac delta in \cref{eq:first_order} requires careful convergence with respect to Brillouin zone sampling . It is common to rewrite \cref{eq:first_order} by neglecting the phonon energy in the delta and by taking the low temperature limit as proposed by Allen.\cite{Allen1972} In most practical applications, vibrational broadening is calculated with this low temperature approximation  \cref{eq:allen}:
\begin{equation}\label{eq:allen}
    \gamma_{\q\nu} \approx 2\pi \hbar\omega_{\q\nu} \sum_{\sigma m n} \int \frac{d\kk}{\Omega_{\mathrm{BZ}}} |g_{mn\nu}(\kk,\q)|^2 \delta(\epsilon_{n\kk}-\epsilon_{F})\delta(\epsilon_{m\kk+\q}-\epsilon_F)
\end{equation}
where $\epsilon_F$ is the Fermi energy. The appendix of Ref.~\cite{Maurer2016a} shows how to arrive at \cref{eq:allen} from \cref{eq:first_order}. \Cref{eq:allen} is positive definite, hence easier to converge numerically than eq.~\ref{eq:first_order}. The disadvantage is that the temperature dependence is lost, and that one cannot resolve fine features on the scale of the phonon energy. Calculations with eq.~\ref{eq:first_order} are somewhat more demanding and have been less commonly reported.~\cite{lazzeriPhononLinewidthsElectronphonon2006,boniniPhononAnharmonicitiesGraphite2007, giustinoElectronphononInteractionUsing2007, Maurer2016a}
For example, Park et al. \cite{Park2007} have used \cref{eq:first_order} to predict the temperature dependence of phonon line broadening in graphite and graphene.

The matrix elements that feature in \cref{eq:first_order} and \cref{eq:allen} and in the phonon self energy expression (\cref{eq:NA_selfenergy}) have units of energy and are defined as follows:
\begin{equation}\label{eq:EPCs_SE}
   g_{mn\nu}(\kk,\q) =  \left(\frac{\hbar}{2M_{\nu}\omega_{\q\nu}}\right)^{1/2} \tilde{g}_{mn\nu}(\kk,\q)
\end{equation}
with 
\begin{equation}\label{eq:EPCs}
    \tilde{g}_{mn\nu} (\kk,\q) = \braket{m\kk+\q|\partial_{\q\nu}V|n\kk}
\end{equation}
The EPC matrix elements in \cref{eq:EPCs} have units of energy per length. They describe the excitation of an electron from a state $n\kk$ to a state $m\kk+\q$ by absorption of a phonon $\q\nu$. $V$ in \cref{eq:EPCs} is the self-consistent ("screened") effective potential from a Kohn-Sham DFT calculation. Its derivative with respect to atomic displacement $\partial_{\q\nu}V$,  is defined as
\begin{equation}\label{eq:perturbation_potential}
    \partial_{\q\nu}V =\sum_{a\kappa} u^{ak}_{\q\nu}  \frac{\partial}{\partial R_{\q, a\kappa}} V(\rr)
\end{equation}
\cref{eq:perturbation_potential} introduces the elements of the phonon displacement eigenmode $u^{a\kappa}_{\q\nu}$ associated with phonon $\q\nu$. Here, $a$ is an atom index and $\kappa$ indicates the 3 Cartesian directions $x$, $y$, and $z$. Note that we do not include mass weighting as the mass weighting is considered in \cref{eq:EPCs_SE} via $M_{\nu}$. By inserting \cref{eq:perturbation_potential} into \cref{eq:EPCs} we find
\begin{equation}\label{eq:EPCs_for_friction}
    \tilde{g}_{mn\nu} (\kk,\q) = \sum_{a\kappa} u^{ak}_{\q\nu}   \braket{m\kk+\q| \frac{\partial}{\partial R_{\q, a\kappa}} V(\rr)  |n\kk} = \mathbf{u}_{\q\nu} \cdot  \tilde{\mathbf{g}}_{mn} (\kk,\q)
\end{equation}
On the right-hand side of \cref{eq:EPCs_for_friction} we introduce a vector notation to denote the multiplication of the normal mode displacement vector $\mathbf{u}_{\q\nu}$ with the vector of atomic contributions to the perturbation potential $\tilde{\mathbf{g}}_{mn} (\kk,\q)$.

%%%%Now we need to define new matrix elements and insert into relaxation rate expression

We can find an expression for the Cartesian friction tensor by inserting  \cref{eq:EPCs_SE} and \cref{eq:EPCs_for_friction} into \cref{eq:first_order}:
\begin{equation}\label{eq:just_before_friction}
\gamma_{\q\nu} = 2\pi\frac{\hbar}{2M_{\nu}\omega_{\q\nu}}\sum_{\sigma mn}\int \frac{d\kk}{\Omega_{BZ}} \left[ \mathbf{u}_{\q\nu} \cdot  \tilde{\mathbf{g}}_{mn} (\kk,\q)   \tilde{\mathbf{g}}_{mn}^{\dagger} (\kk,\q) \cdot \mathbf{u}^T_{\q\nu} \right] (f_{n\kk}-f_{m\kk+\q}) \delta(\epsilon_{m\kk+\q}-\epsilon_{n\kk}-\hbar\omega_{\q\nu})
\end{equation}
The factor of 2 in this expression arises from the definition of FWHM in \cref{eq:ep-broadening} and is cancelled by the factor 2 in the definition of the EPC matrix elements.~\cite{calandraAdiabaticNonadiabaticPhonon2010}

We can bring the normal mode displacement vectors $\mathbf{u}_{\q\nu}$ outside of the sum over states and define the mass-weighted normal mode displacement vectors $\tilde{\mathbf{u}}_{\q\nu}=\mathbf{u}_{\q\nu}/\sqrt{M_{\nu}}$
\begin{align}\label{eq:broadening_with_friction}
    \gamma_{\q\nu} = \hbar \Gamma_{\q\nu} = \hbar \left[ \tilde{\mathbf{u}}_{\q\nu} \mathbf{\Lambda}^{\q}(\hbar\omega_{\q\nu})\tilde{\mathbf{u}}^T_{\q\nu}  \right]
\end{align}
Herein, we have defined the Cartesian friction tensor\cite{Maurer2016a}
\begin{equation}\label{eq:friction_tensor_single_delta}
    \Lambda_{a\kappa,a'\kappa'}^{\q,\mathrm{I}}(\hbar\omega) = \pi\hbar \sum_{\sigma mn}\int \frac{d\kk}{\Omega_{BZ}}  \tilde{g}^{a\kappa}_{mn} (\kk,\q)   (\tilde{g}^{a'\kappa'}_{mn})^{*} (\kk,\q)   (f_{n\kk}-f_{m\kk+\q}) \frac{\delta(\epsilon_{m\kk+\q}-\epsilon_{n\kk}-\hbar\omega)}{\hbar\omega}
\end{equation}
A similar result for the phonon linewidth can be generated on the basis of atomic response correlation functions. \cite{calandraAdiabaticNonadiabaticPhonon2010, Head-Gordon1995}

We can follow a similar derivation starting from the low temperature double delta expression by Allen in \cref{eq:allen} and arrive at:
\begin{equation}\label{eq:friction_tensor_double_delta}
    \Lambda_{a\kappa,a'\kappa'}^{\q,\mathrm{II}} = \pi\hbar \sum_{\sigma mn}\int \frac{d\kk}{\Omega_{BZ}}  \tilde{g}^{a\kappa}_{mn} (\kk,\q)   (\tilde{g}^{a'\kappa'}_{mn})^{*} (\kk,\q)\delta(\epsilon_{n\kk}-\epsilon_F) \delta(\epsilon_{m\kk+\q}-\epsilon_F) 
\end{equation}

Both these expressions of electronic friction are generalizations of the expressions defined by Maurer et al.\cite{Maurer2016a} for arbitrary $\q$ values and  both have units of [mass/time] or [(energy* time)/length$^2$] or [action/length$^2$].

\subsection{EPC matrix elements and expressions in local atomic orbital basis}
% ---------------------------------------------------------------- %

Several previous works have discussed the implementation of EPCs in local orbital basis. \cite{Savrasov1996, Gunst2016,Maurer2016a, agapitoInitioElectronphononInteractions2018} We will provide a brief summary of the main expressions below. We can find expressions for the EPCs $\tilde{g}^{a\kappa}_{mn} (\kk,\q)$ by expanding the electronic eigenstates $m\kk$ in a local orbital basis of crystal periodic atomic orbitals $\phi_{i\kk}$:
\begin{equation}\label{eq:lcao}
    \ket{m\kk} = \psi_{m\kk} = \sum_i C^i_{m\kk}  \phi_{i\kk}(\rr) .
\end{equation}

Inserting \cref{eq:lcao} into \cref{eq:EPCs} yields
\begin{equation}\label{eq:local_EPC1}
    \tilde{g}^{a\kappa}_{mn} (\kk,\q) = \sum_{ij}  \left(C^{j}_{m\kk+\q}\right)^* C^{i}_{n\kk}  \underbrace{\braket{\phi_{i\kk+\q} | \frac{\partial  }{\partial R_{\q, a\kappa}}V(\rr)|\phi_{j\kk}}}_{=h^{a\kappa}_{ij}(\kk,\q)}
\end{equation}

We start with $h_{ij}^{a\kappa}(\kk,\q)$ defined in \cref{eq:local_EPC1} and simplify notation. We first recognize that the perturbation of the effective potential is equivalent to the Hamiltonian response.
\begin{equation}
 h_{ij}^{a\kappa}(\kk,\q) = \braket{\phi_{i\kk+\q} | \frac{\partial  }{\partial R_{\q, a\kappa}}V|\phi_{j\kk}} = \braket{\phi_{i\kk+\q} | \frac{\partial  }{\partial R_{\q, a\kappa}}H|\phi_{j\kk}}
\end{equation}
The matrix elements $h^{a\kappa}_{ij}(\kk,\q)$ define a first-order Hamiltonian response matrix $H^{(1)}$ that can be calculated in FHI-aims via Density Functional Perturbation Theory (DFPT)\cite{shangLatticeDynamicsCalculations2017} or finite difference response \cite{Maurer2016a}. Here in this work, we will only consider finite-difference-based calculations. While $h_{ij}^{a\kappa}(\kk,\q)$ can be directly calculated with DFPT, in the case of finite difference evaluation of the response with consecutive self-consistent-field calculations for displaced atoms, we only have access to 
\begin{equation}
H_{ij}^{a\kappa,(1)}(\kk,\q) = \frac{\partial }{\partial R_{\q, a\kappa}}\braket{\phi_{i\kk+\q}|H|\phi_{j\kk}}
\end{equation}
As previously shown, following relationship between the derivative of the expectation value and the derivative of the operator holds: \cite{Head-Gordon1995,Savrasov1996, Maurer2016a}
\begin{equation}\label{eq:braket_deriv}
   \braket{\phi_{i\kk+\q}|\frac{\partial  }{\partial R_{\q, a\kappa}}H|\phi_{j\kk}} = H_{ij}^{a\kappa,(1)}(\kk,\q)  - \braket{\frac{\partial \phi_{i\kk+\q}}{\partial R_{\q, a\kappa}}|H|\phi_{j\kk}} - \braket{\phi_{i\kk+\q}|H|\frac{\partial \phi_{j\kk} }{\partial R_{\q, a\kappa}}}
\end{equation}

We can now insert \cref{eq:braket_deriv} into \cref{eq:local_EPC1}:
\begin{equation}
    \tilde{g}^{a\kappa}_{mn} (\kk,\q) = \sum_{ij}  \left(C^{j}_{m\kk+\q}\right)^*  \left(H_{ij}^{a\kappa,(1)}(\kk,\q) - \braket{\frac{\partial \phi_{i\kk+\q}}{\partial R_{\q, a\kappa}}|H|\phi_{j\kk}} - \braket{\phi_{i\kk+\q}|H|\frac{\partial \phi_{j\kk} }{\partial R_{\q, a\kappa}}} \right)  C^{i}_{n\kk} 
\end{equation}
Using the fact that the wavefunctions satisfy a generalized eigenvalue equation, we can  make following replacement:
\begin{equation}
    \tilde{g}^{a\kappa}_{mn} (\kk,\q) =   \sum_{ij}  \left(C^{j}_{m\kk+\q}\right)^*   \left(H_{ij}^{a\kappa,(1)}(\kk,\q) - \epsilon_{n\kk}\braket{\frac{\partial \phi_{i\kk+\q}}{\partial R_{\q, a\kappa}}|\phi_{j\kk}} - \epsilon_{m\kk+\q}\braket{\phi_{i\kk+\q}|\frac{\partial \phi_{j\kk} }{\partial R_{\q, a\kappa}}} \right)  C^{i}_{n\kk}     
\end{equation}
The second and third term on the right hand side correspond to the left and right sided derivative components of the overlap matrix  $S_{ij, \kk}^{a\kappa, L}$ and $S_{ij, \kk}^{a\kappa, R}$:
\begin{equation}
    S_{ij}^{a\kappa,(1)}(\kk,\q) = \frac{\partial}{\partial R_{\q,a\kappa}}\braket{\phi_{i\kk+\q}|\phi_{j\kk}} = \braket{\frac{\partial \phi_{i\kk+\q}}{\partial R_{\q, a\kappa}}|\phi_{j\kk}} + \braket{\phi_{i\kk+\q}|\frac{\partial \phi_{j\kk} }{\partial R_{\q, a\kappa}}} =  S_{ij}^{a\kappa,L}(\kk,\q)  + S_{ij}^{a\kappa,R}(\kk,\q) 
\end{equation}
With this, we come to following expression of the EPC matrix elements of \cref{eq:local_EPC1}:
\begin{equation}\label{eq:local_EPC1_v2}
    \tilde{g}^{a\kappa}_{mn} (\kk,\q) = \sum_{ij}  \left(C^{j}_{m\kk+\q}\right)^* C^{i}_{n\kk}  \left( H_{ij}^{a\kappa,(1)}(\kk,\q) - \epsilon_{n\kk}S_{ij}^{a\kappa,L}(\kk,\q) - \epsilon_{m\kk+\q}S_{ij}^{a\kappa,R}(\kk,\q)  \right)  
\end{equation}

Comparing with the findings of Maurer et al., \cite{Maurer2016a} we can relate the potential derivative (deformation potential matrix elements) and nonadiabatic coupling matrix elements as follows:
\begin{equation}
    \braket{\psi_{m\kk+\q}|\frac{\partial V}{\partial R_{\q, a\kappa}}|\psi_{n,\kk}} =  (\epsilon_{m\kk+\q}-\epsilon_{n\kk})\cdot \braket{\psi_{m\kk+\q}|\frac{\partial}{\partial R_{\q, a\kappa}}|\psi_{n,\kk}}
\end{equation}

As shorthand, we define a EPC coupling matrix $G_{ij}^{a\kappa,(1)}(\kk,\q)$ in local orbital representation as
\begin{equation}\label{eq:EPC_matrix1}
\mathbf{G}^{nm,(1)}_{ij}(\kk,\q) = \left( \mathbf{H}_{ij}^{(1)}(\kk,\q) - \epsilon_{n\kk}\mathbf{S}_{ij}^{L}(\kk,\q) - \epsilon_{m\kk+\q}\mathbf{S}_{ij}^{R}(\kk,\q)  \right)
\end{equation}

In previous works, for computational convenience, approximate versions of these matrix elements have been proposed by Head-Gordon and Tully \cite{Head-Gordon1995, Maurer2016a}, where the dependence on individual electronic states is dropped and the full linear response of the overlap matrix is used instead:
\begin{equation}\label{eq:EPC_matrix_HGT}
\mathbf{G}_{ij}^{\mathrm{HGT}}(\kk,\q) = \left( \mathbf{H}_{ij}^{(1)}(\kk,\q) - \epsilon_F\mathbf{S}_{ij}^{(1)}(\kk,\q)  \right)
\end{equation}
Maurer et al. have assessed the impact of this approximation on vibrational relaxation rates for aperiodic systems and have found it to affect final predictions by only few percentage points.~\cite{Maurer2016a} The effect of this approximation on periodic calculations has not been assessed before and is discussed in section \ref{sec:convergence}.

\subsection{Delta function approximations}
% ---------------------------------------------------------------- %
The numerical evaluation of the friction tensor involves either two delta functions (low temperature limit, \cref{eq:friction_tensor_double_delta}) or a single delta function (\cref{eq:friction_tensor_single_delta}). These delta functions cannot be reliably evaluated directly for ab initio calculations due to the discrete nature of the density of states represented on a finite $\kk$-point mesh. Typically the delta function(s) are replaced by a broadening function of finite width, common choices include Gaussian and Lorentzian functions. 
The most common choice is a Gaussian function:
%Momentum indices?
\begin{equation}\label{eq:Gaussian}
    \hat{\delta}\left(\epsilon_{i}-\epsilon_{j}\right)=\frac{1}{\sqrt{2 \pi} \sigma} \exp \left\{\frac{-\left(\epsilon_{i}-\epsilon_{j}\right)^{2}}{2 \sigma^{2}}\right\}
\end{equation}

An additional normalization technique was previously proposed for single delta function calculations when relatively large smearing widths are applied \cite{Maurer2016a}. This is to account for the contribution of the Gaussian that falls below the integration window:
\begin{equation}\label{eq:normalised_Gaussian}
    \tilde{\delta}\left(\epsilon_{i}-\epsilon_{j}\right)=\frac{\hat{\delta}\left(\epsilon_{i}-\epsilon_{j}\right)}{\int_{0}^{\infty} \hat{\delta}\left[\epsilon-\left(\epsilon_{i}-\epsilon_{j}\right)\right] d \epsilon}=\frac{\hat{\delta}\left(\epsilon_{i}-\epsilon_{j}\right)}{\frac{1}{2}\left[1-\operatorname{erf}\left(\frac{\epsilon_{j}-\epsilon_{i}}{\sqrt{2}\sigma}\right)\right]}
\end{equation}
We will reassess this approach in section \ref{sec:convergence}. For double delta functions the effective smearing is twice that of a single delta function, thus we define an effective smearing parameter, $\sigma_\mathrm{eff} = N_\mathrm{delta} \times \sigma$ to be able to compare the two. A Lorentzian broadening function can be defined as follows:
%Lorentzian
\begin{equation}
    \hat{\delta}_\mathrm{Lorentz.}\left(\epsilon_{i}-\epsilon_{j}\right)=\frac{1}{\pi} \frac{\frac{1}{2} \sigma}{\left(\epsilon_{i}-\epsilon_{j}\right)^{2}+\left(\frac{1}{2} \sigma\right)^{2}}
\end{equation}

%Methfessel-paxton [PRB 1989]
%We are a type III integrand

%Adaptive broadening
%J. R. Yates, X. Wang, D. Vanderbilt, and Souza I. Phys. Rev. B, 75:195121, 2007.
%Wannier integration only? need to calculate eigenvalue spacing , so they calc the derivative of the eigenvalues on a uniform k grid
% ================================================================ %
   \section{Methods}
% ================================================================ %

\subsection{Implementation}

\begin{figure}
    \centering
    \includegraphics[width=2.in]{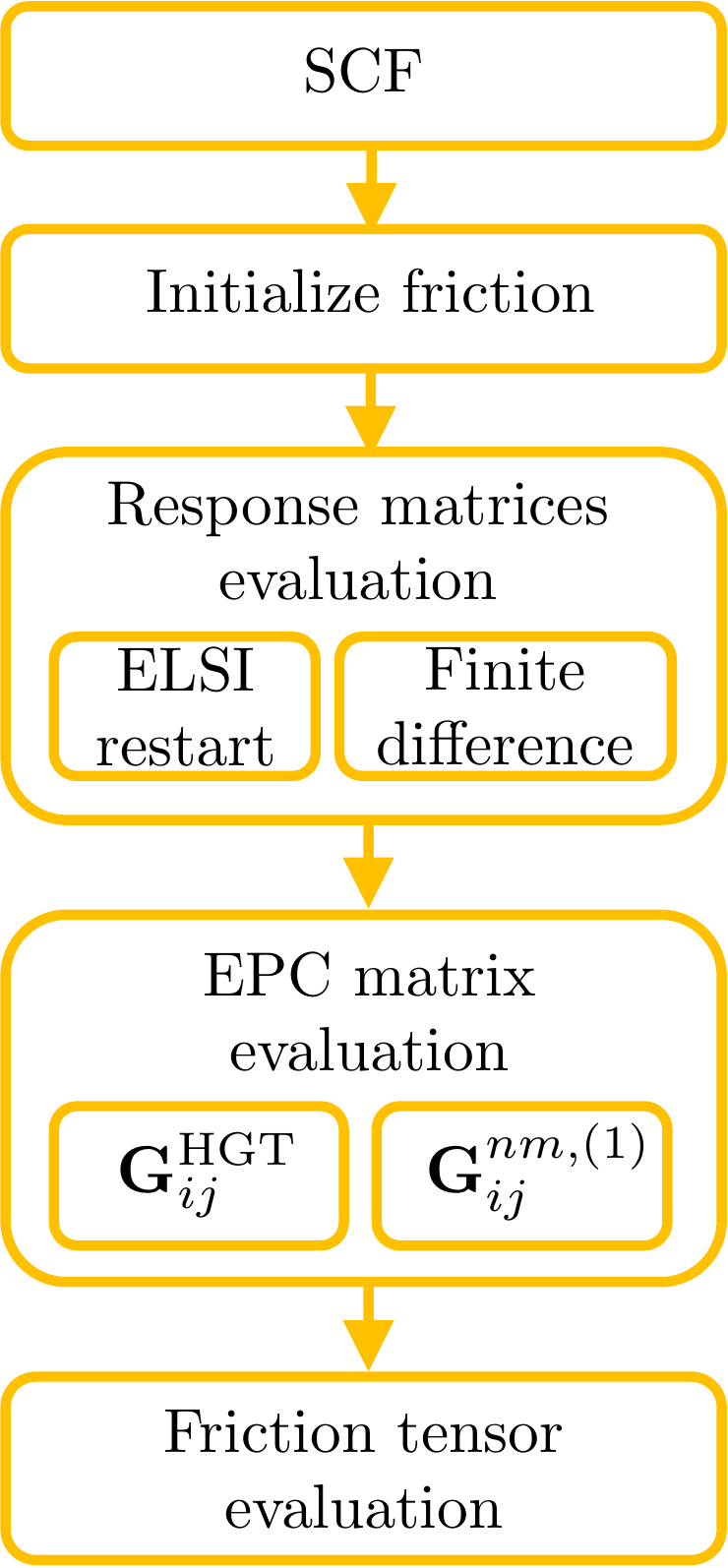}
    \caption{Sequence of calculation operations required during the calculation of the electronic friction tensor.}
    \label{fig:code_layout}
\end{figure}

The workflow for the evaluation of the friction tensor as implemented in FHI-aims is shown in  \cref{fig:code_layout}. The subset of atoms to be included in the electronic friction calculation can be specified by introducing the keyword \texttt{friction\_atom} after the definition of the Cartesian position of the atom in the geometry.in input file. This means that the response for this atom in each Cartesian direction will be calculated. 

The electronic friction calculation begins with a ground state self-consistent-field (SCF) DFT calculation until convergence is reached, this is required as the ground state eigenvectors and chemical potential (Fermi energy) are used in the calculation of the response. The calculation of the required response matrices (typically the first-order Hamiltonian and overlap matrices) follows. These can be read in from a prior calculation using input and output routines provided via the ELSI library, or calculated using finite difference. The latter requires displacement of each friction atom, forwards and backwards in each Cartesian direction, and the SCF to be recalculated in each case. Since the whole process takes place in a single calculation, the ground state wavefunction can be used as an initial guess for each finite difference SCF, facilitating convergence. If the analytical matrix elements $\mathbf{G}^{nm,(1)}_{ij}$ (\cref{eq:EPC_matrix1}) are required rather than the approximate $\mathbf{G}_{ij}^{\mathrm{HGT}}$ (\cref{eq:EPC_matrix_HGT}), a calculation of the one-sided derivative of the overlap matrix follows. Now the coupling matrix can be obtained, the type depends on the choice of non-adiabatic coupling matrix elements. Finally, the tensor is evaluated and printed. During the tensor evaluation, several calculation parameters can be selected. This includes the value of $\sigma$ for the broadening of the delta function, the value of the temperature that defines the orbital occupations, the choice of delta function, the type of friction tensor expression to be used (\cref{eq:friction_tensor_single_delta} or \cref{eq:friction_tensor_double_delta}), and the frequency at which the friction tensor is evaluated (if  \cref{eq:friction_tensor_single_delta} is used). The computational performance and scalability is of the code is described in section \ref{sec:scaling}.

\subsection{\label{sec:scaling}Code Scalability and Performance}
% ---------------------------------------------------------------- %

\begin{figure}
    \centering
    \includegraphics{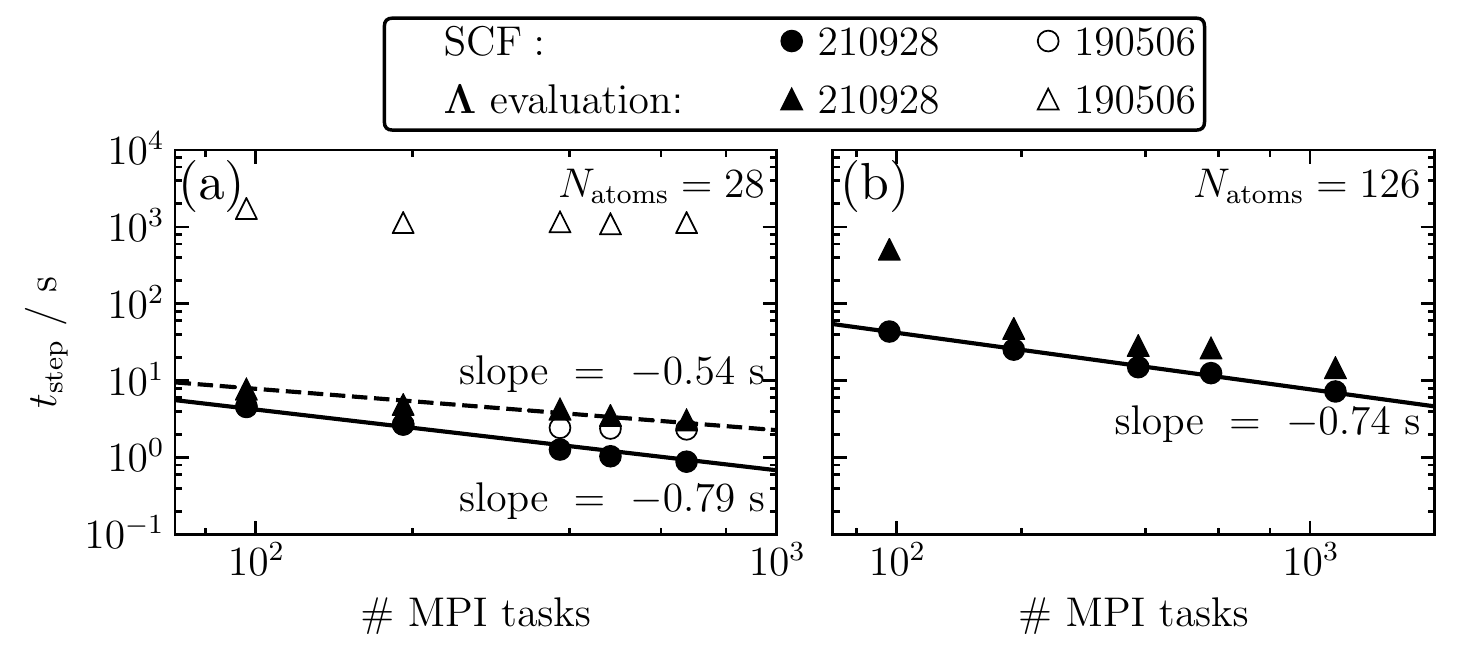}
    \caption{Scaling behaviour for the electronic friction module using finite difference linear response calculations for CO/Cu(100). We compare each SCF step and the friction tensor ($\bm{\Lambda}$ evaluation, one per calculation). In (a) we compare the older code (190506, open symbols) with the current refactored code (210928, closed symbols) (256 $\kk$-points). In (b) we use a larger system (64 $\kk$-points), which was not feasible with the original code. Calculations performed on Warwick SC RTP Avon HPC. Raw data used to produce this figure is available in the NOMAD repository under: URL to be added after acceptance.}
    \label{fig:scalability}
\end{figure}

The electronic friction code has been substantially refactored since it was originally published. \cite{Maurer2016a} The response matrices now support ScaLAPACK-type matrix distribution, allowing efficient distribution and scaling when $N_\mathrm{cores}>N_k$. This allows much larger systems to be investigated, the largest system calculated in this work comprises 252 atoms, 16 $\kk$-points and 108 response coordinates (3 Cartesian coordinates per 36 response atoms). The original code would not be able to treat this system, or any system with more than approximately 10 atoms for which the response is calculated. In \cref{fig:scalability}(a) the improved scaling behaviour of the refactored code (version 210928) is compared to the original code (version 190506). The refactored code significantly outperforms the original code even when $N_\mathrm{cores}<N_k$, specifically for the friction tensor evaluation. The scaling of the original code saturates with cores when $N_\mathrm{cores}=N_k$, whilst the refactored code continues to scale beyond this point. The scaling remains favourable for larger system that were not feasible with original code, as shown in \cref{fig:scalability}(b). The friction tensor evaluation remains on a similar timescale to a single SCF iteration, and shows similar scaling with the number of cores. Key benefits of having an internal finite difference code (rather than calling FHI-aims externally multiple times via post-processing) is that i) the ground-state wavefunction can be used as an initial guess for finite-difference displacements to reduce SCF iterations for all finite difference displacements, ii) time consuming I/O is kept to a minimum iii) evaluation of the friction tensor runs internally in parallel. Whilst the full calculation of the friction tensor is still too computationally costly for large-scale on-the-fly ab initio MDEF simulations, the code improvements will dramatically enhance our ability to gather data for the construction of machine learning models of the electronic friction tensor and the electron-phonon response.\cite{Zhang2020} 

% mention that frozen core works with friction?
% Scaling figure:

\subsection{Computational Details}

%FHIaims
Unless stated otherwise, all calculations are performed using the all-electron NAO code FHI-aims (2021 development version) \cite{Blum2009} without spin polarization and with PBE exchange-correlation functional, \cite{Perdew1996} dipole correction, atomic ZORA relativistic correction,  tolerances of $10^{-6}$~eV, $10^{-3}$~eV and $10^{-5}$~$\mathrm{e}$~/~$\mathrm{a_0}^3$ for the total energy, the eigenvalue energies and the electronic density respectively. Unless stated otherwise, calculations are performed with standard tight basis set definitions. All surface slabs feature a vacuum slab of at least 50 $\mathrm{\AA{}}$ in $z$ direction. Surface slabs are constructed with PBE optimized bulk lattice constants. Unless stated otherwise, all friction tensors were evaluated using finite difference (displacement of 0.001 $\mathrm{\AA{}}$ with a Fermi factor corresponding to an electronic temperature of 300~K, a value of 0.6~eV for $\sigma$ for a normalized Gaussian that approximates the single delta function (\cref{eq:normalised_Gaussian}). Unless exploring convergence with respect to substrate thickness (where only the largest slab was relaxed and  others were generated from that slab), structures were relaxed using BFGS algorithm with a tolerance of 0.01~$\mathrm{eV}/\mathrm{\AA}$ and the bottom 2 substrate layers were constrained.

% SECTION 4.2
%QE
For the benchmarking of the code (Table \ref{tab:coru} and \ref{tab:h2cu}), we have aimed to reproduce the results given in Ref. \cite{spiering_testing_2018} and Ref. \cite{diesen_ultrafast_2021} with {\sc Quantum ESPRESSO} (version 6.8) \cite{Giannozzi2009} and FHI Aims (2021 developmental version) codes. Our settings are chosen to be as close as possible to the settings discussed in the two literature studies. In short, for H$_2$/Cu(111) system, we employ PW91 \cite{perdew_atoms_1992} exchange-correlation functional with ONCV pseudopotentials \cite{hamann_optimized_2013}, plane-wave energy cutoff of 816~eV and 18$\times$18$\times$1 $\kk$-point grids with p(2$\times$2) Cu(111) slabs containing 4 layers of metal and a vacuum height of 10 \AA. The settings used for CO/Ru(0001) calculations are equivalent to those used by Diesen et al \cite{diesen_ultrafast_2021}; RPBE \cite{hammer_improved_1999} exchange-correlation functional with ultrasoft pseudopotentials \cite{Vanderbilt1990}, plane-wave energy cutoff of 544~eV for p(1$\times$1) and p(2$\times$2) Ru(0001) slabs (1~ML and 0.25~ML coverage respectively), containing 6 layers of metal and a vacuum of 20 \AA. The 96$\times$96$\times$1 and 48$\times$48$\times$1 $\kk$-point grid was used respectively. For both systems we utilized the $\kk$-grid interpolation scheme as employed in the PHonon QE module. The delta function was approximated with a standard Gaussian broadening (\cref{eq:Gaussian}) of 0.6~eV in both systems.
%explain settings for H2/Cu and CO on Ru calculations here.
We employed the identical settings in FHI-aims, except using NAO basis sets: 'tight' for calculating forces and 'light' for calculating electronic friction. Additionally, for calculations with H$_2$/Cu(111) system we use 16$\times$16$\times$1 $\kk$-point grids.
To obtain relaxed structures for the layer convergence tests of CO/Ru, in both codes, we removed the bottom layers of the optimized CO/Ru(0001) structure containing 16 layers of Ru and decreased the vacuum height accordingly. To calculate electronic friction for the H$_2$/Cu(111) system with 4 and 6 layers, we optimized the H$_2$/Cu structures with 4 layers and located the transition state with the nudged elastic band (NEB) method.\cite{Henkelman2000a} Subsequently we added two additional bottom layers to extend the slab to 6 layers.

%NOMAD DOI
Each figure and table caption contains a DOI that links to a dataset in the NOMAD materials repository where all input and output files are available for download.

% ================================================================ %
    \section{Results and Discussion}
% ================================================================ %

\subsection{\label{sec:convergence}Convergence of electronic friction tensor and vibrational linewidth calculations}
% ---------------------------------------------------------------- %

\begin{figure}
    \centering
    \includegraphics{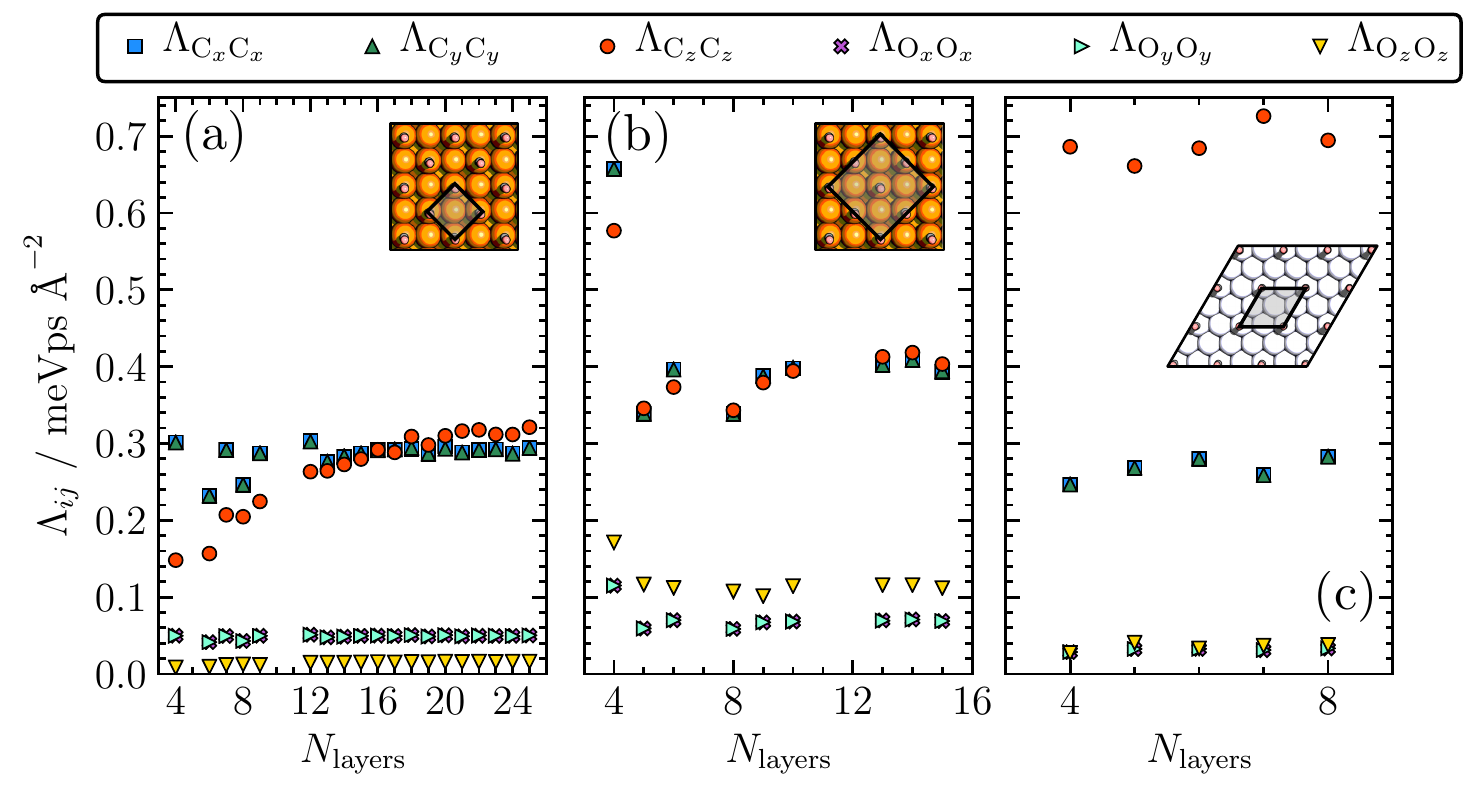}
    \caption{Convergence of the diagonal elements of $\bm{\Lambda}$ as a function of the number of substrate layers for 3 systems. CO/Cu(100) with a c($2\times2$) coverage is shown in (a) and (b) for a primitive and supercell, respectively, whilst a p($2\times2$) coverage of CO/Pt(111) is shown in (c).  Both CO/Cu(100) cells have 6 substrate layers, whilst CO/Pt(111) has been calculated with 4 layers. The calculations here employ the delta function approximation given in \cref{eq:normalised_Gaussian} with $\sigma=0.6$~eV. Insets show a graphical depiction of the unit cells. Raw data used to produce this figure is available in the NOMAD repository under: URL to be added after acceptance.}
    \label{fig:tensor_layers_all}
\end{figure}

\begin{figure}
    \centering
    \includegraphics[]{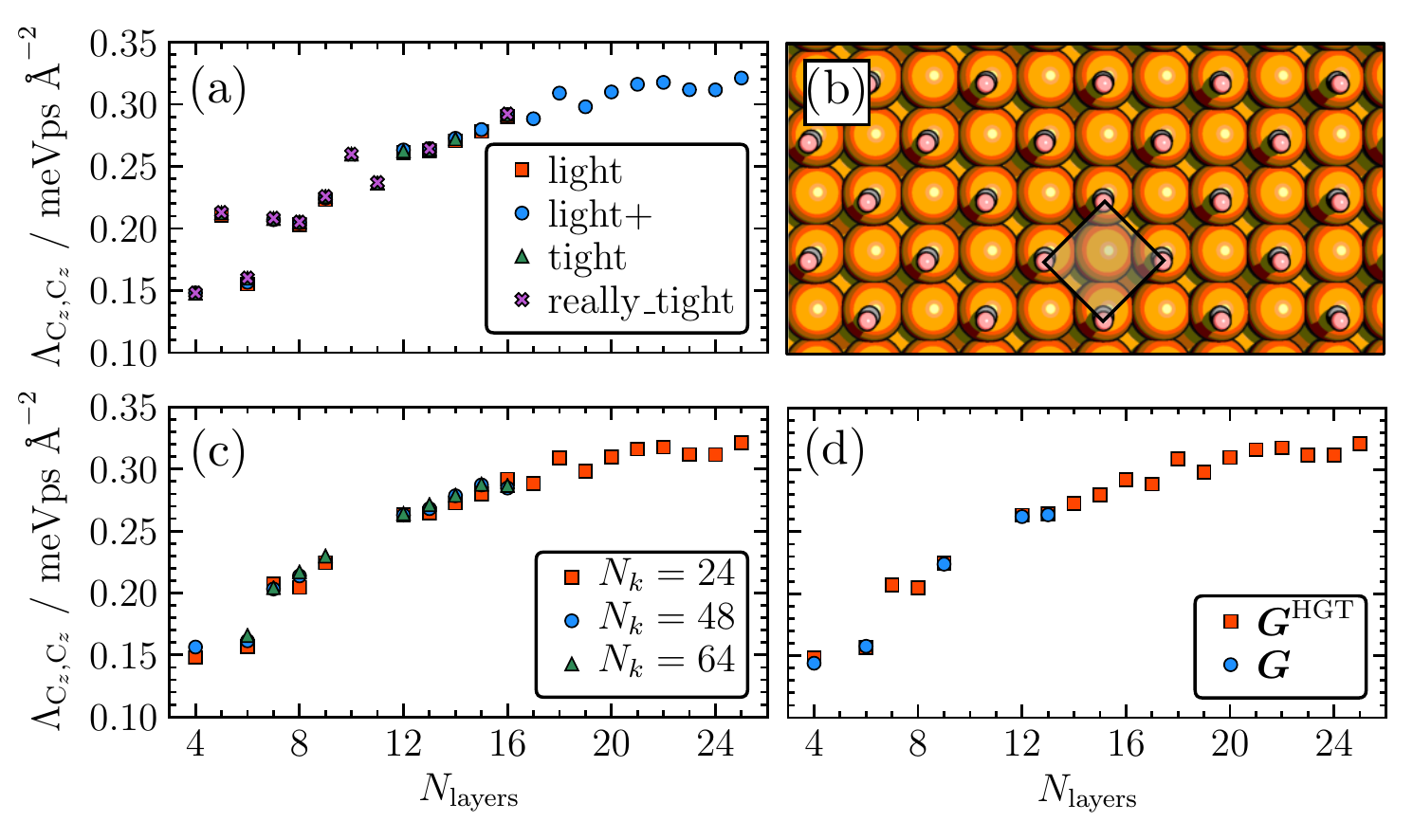}
    \caption{Convergence of $\Lambda_{\mathrm{C}_z\mathrm{C}_z}$ component of the electronic friction tensor for primitive c($2\times2$) CO/Cu(100) with respect to number of substrate layers ($x$-axis), basis set size (a) and $\kk$-grid size (c). Comparison of HGT elements and analytical elements shown in (d). The calculations here employ the delta function approximation given in \cref{eq:normalised_Gaussian} with $\sigma=0.6$~eV. Graphical depiction of cell is given in (b). Raw data used to produce this figure is available in the NOMAD repository under: URL to be added after acceptance.}
    \label{fig:co_cu_prim_basis_layers}
\end{figure}

\begin{figure}
    \centering
    \includegraphics{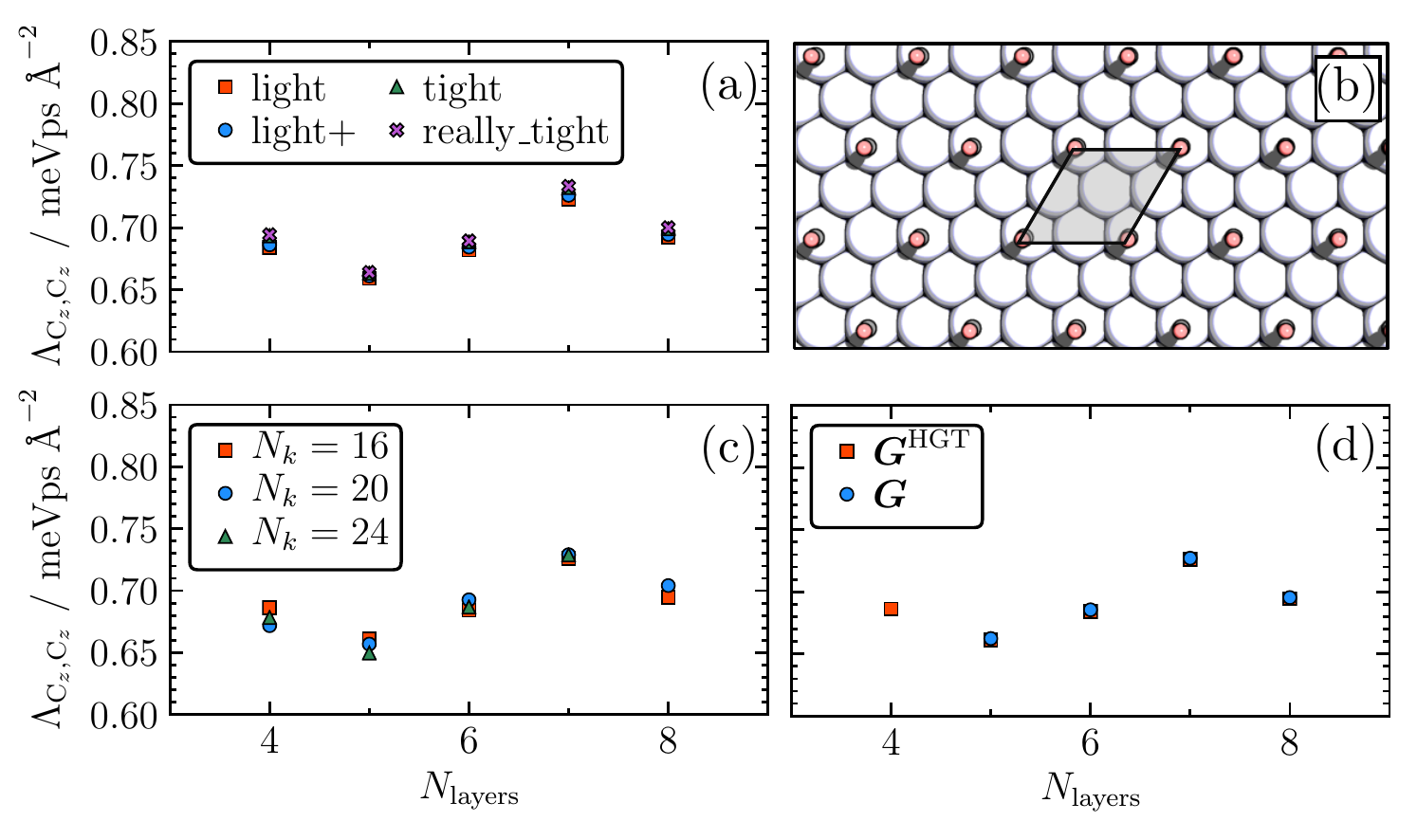}
    \caption{Same as \cref{fig:co_cu_prim_basis_layers} except for p($2\times2$) CO/Pt(111). Raw data used to produce this figure is available in the NOMAD repository under: URL to be added after acceptance.}
    \label{fig:co_pt_convergence}
\end{figure}

% Par 1 - All tensor elements for all 3 systems
We first explore the convergence of the computed friction tensor as defined in \cref{eq:friction_tensor_single_delta} using FHI-aims with respect to several calculation parameters for two exemplary systems, namely a c($2\times2$) overlayer of CO molecules adsorbed at Cu(100) and a p($2\times2$) overlayer of CO adsorbed at Pt(111). The overlayers are depicted in the insets of \cref{fig:tensor_layers_all}(a) and (c) (also later in \cref{fig:co_cu_prim_basis_layers}(b) and \cref{fig:co_pt_convergence}(b)). The two systems are known to display significantly different convergence behaviour, which is related to differences in projected density-of-state (DOS) and their band structure. We also compare the primitive copper cell to a supercell of the same coverage, containing 4 CO molecules (depicted in \cref{fig:tensor_layers_all}(b)). 

The diagonal electronic friction tensor elements, $\Lambda_{ii}$,  show different convergence behaviour with respect to the number of included substrate layers. For the primitive copper cell, the $\Lambda_{\mathrm{C}_z\mathrm{C}_z}$ component of the friction tensor is the last to reach convergence to within $\pm$0.01~$\mathrm{meV}\mathrm{ps}\mathrm{\AA}^{-2}$, at about 20 substrate layers. This element exhibits a general monotonic increase with the number of substrate layers that is not seen for any other diagonal element for the Cu supercell nor the Pt substrate. This behaviour may be attributed to the slow embedding of the surface states into bulk as the substrate gets thicker. \cite{Kratzer2019} The other diagonal C components converge quicker to similar accuracy at approximately 12 substrate layers. The oxygen centric diagonal components appear well converged even at 4 layers. The convergence behaviour is significantly changed for $\Lambda_{\mathrm{C}_z\mathrm{C}_z}$ in the supercell, where it displays similar behaviour to the other C diagonal components. Here, 6 layers is sufficient to converge all components within $\pm$0.06~$\mathrm{meV}\mathrm{ps}\mathrm{\AA}^{-2}$. The p($2\times2$) cell of CO/Pt(111) exhibits well converged behaviour for all elements at just 4 layers within a tolerance of 10~\%. In all cases the C-based diagonal elements exhibit larger magnitudes due to the proximity of the C atom to the surface compared to oxygen. The rest of this section will focus on the convergence behaviour for $\Lambda_{\mathrm{C}_z\mathrm{C}_z}$ for all 3 systems, since it shows the slowest convergence for the primitive copper cell.

% Par 2 - basis, kgrid, HGT element for prim CO/Cu(100)

\cref{fig:co_cu_prim_basis_layers}(a) shows that the basis set convergence of electronic friction tensor calculations is quickly achieved with the standard NAO basis set definitions, even at the "light" default set of basis definitions. \cite{Blum2009} The friction tensor elements did not differ when varying the confinement radius of the basis functions. No difference was found between a $3$--$6\ \mathrm{\AA{}}$ onset radius of the cutoff potential for the 6 layer cell (not shown). Change of the basis set does not affect the convergence with respect to the number of substrate layers. \Cref{fig:co_cu_prim_basis_layers}(b) shows that this also holds for changes in the Monkhorst-Pack $\kk$-grid size \cite{Monkhorst1976}, as long as the $\kk$-grid is above a sufficiently converged threshold of $N_k\times  N_k\times 1$, where $N_k$ is above 24. Finally, in \cref{fig:co_cu_prim_basis_layers}d we compare the effect of using the approximate EPC matrix elements as defined by Head-Gordon and Tully, $\mathbf{G}_{ij}^{\mathrm{HGT}}$ against using the exact matrix elements, $\mathbf{G}^{nm,(1)}_{ij}$. As can be seen, the effect of this approximation is negligible and within the numerical tolerance of the calculations. This suggests that this approximation is justified for periodic systems and aperiodic calculations on metal cluster cut-outs as previously assessed by Maurer et al.\cite{Maurer2016a}.

%Primitive

\begin{figure}
    \centering
    \includegraphics{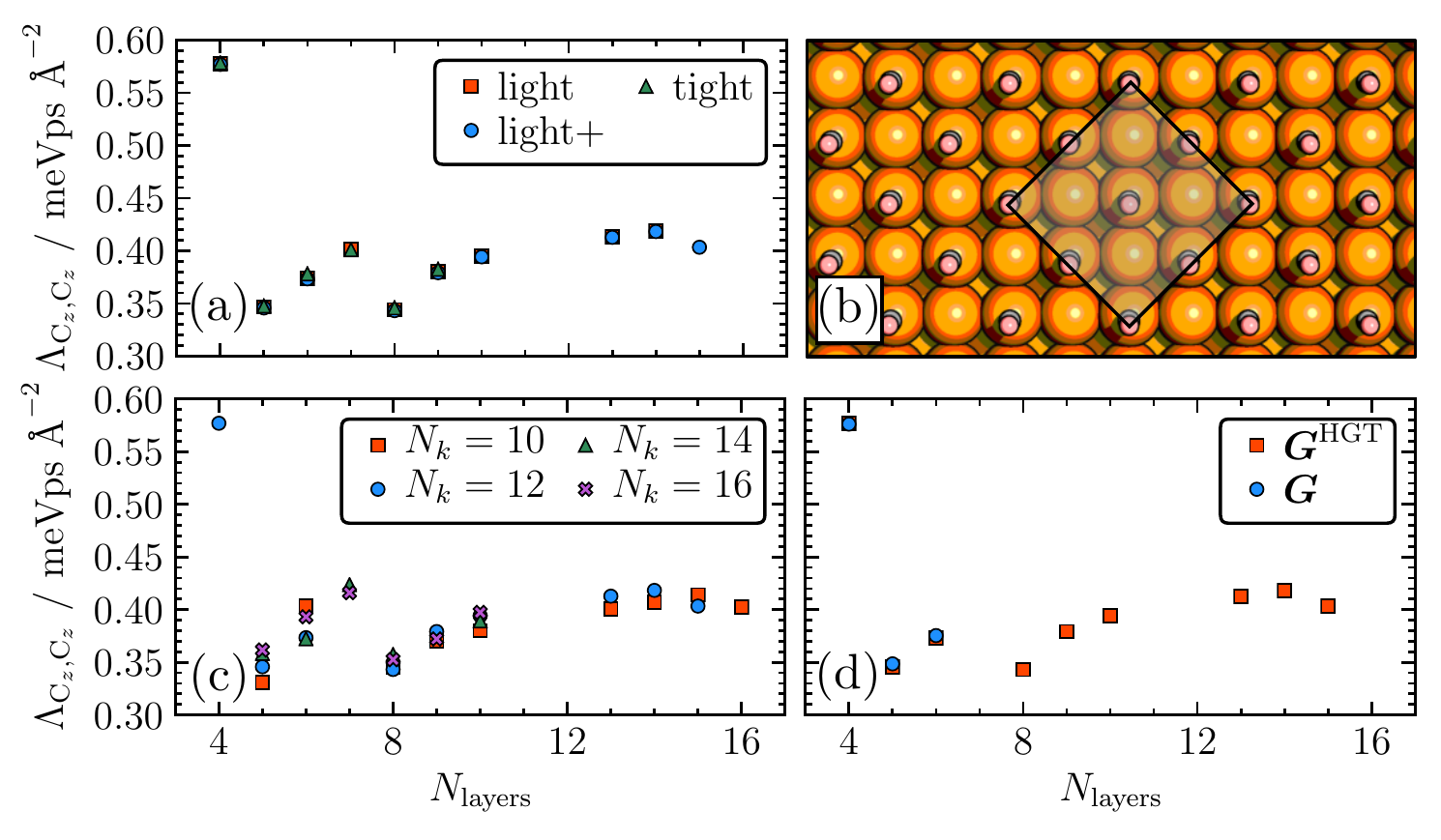}
    \caption{Same as \cref{fig:co_cu_prim_basis_layers} except for a supercell. Raw data used to produce this figure is available in the NOMAD repository under: URL to be added after acceptance.}
    \label{fig:co_cu_supercell_convergence}
\end{figure}

% Par 3 - basis, kgrid, HGT element for bigger super CO/Cu(100)
By doubling the supercell in which we represent the c($2\times2$) overlayer in both lateral directions (\cref{fig:co_cu_supercell_convergence}(b)), we find that the convergence with respect to the number of substrate layers is accelerated as can be seen in panel a, c, and d of the same figure. The change in friction matrix element upon adding a substrate layer is already within 10 \% of the total value after only 5 layers. Whilst the basis is similarly well converged (panel a), even at light settings, and the HGT elements do not differ significantly from the exact ones (panel d), there is more prominent instability when changing $\kk$-grid size compared to the primitive cell (panel c). Nonetheless, the friction tensor values are converged with respect to $\kk$-grid with a $12\times12\times1$ $\kk$-grid (equivalent to $24\times24\times1$ in the primitive unit cell) within $\approx 5 \%$ of larger $\kk$-grid values. Note that the friction tensor element converges to a different value than in the case of the primitive unit cell (0.40 and 0.32 $\mathrm{meV}\mathrm{ps}\mathrm{\AA}^{-2}$, respectively), which is due to the fact that the increase of the unit cell leads to band folding into a reciprocal unit cell of half the size. The result is that nominal intraband excitations can be described by momentum-conserving interband excitation in this larger unit cell, which changes the spectral density. The higher density of available excitations is likely also what leads to the improved convergence with respect to the number of substrate layers. The inclusion of nominal intraband excitations will be discussed in more detail in section \ref{sec:intraband}.

%Supercell

% Par 4 - - basis, kgrid, HGT element for CO/Pt(111)
The convergence behaviour of EPC linewidths and electronic friction tensor elements is strongly dependent on the metal slab and system under study and Cu(100) is a notoriously difficult case. The convergence of the $\Lambda_{\mathrm{C}_z\mathrm{C}_z}$ component of CO adsorbed on Pt(111) as shown in \cref{fig:co_pt_convergence} is rapidly converged with the number of layers, within $\pm$ 10\% with just 4 substrate layers. Despite also being a primitive cell, the friction tensor evaluation is stable with respect to the number of substrate layers. The change in the friction tensor element is slight for the explored basis set and $\kk$-grid size parameters, with minor differences only visible when using 4 substrate layers. Again, there is no significant observed difference between the HGT and analytical elements.

% CO/Pt(111)

% Par 5 - briefly summarize layer dependence and relate to literature
In summary, we find that the convergence with respect to numerical model parameters is fairly swift with the exception of the number of substrate layers that must be accounted for. It is generally known that some properties are particularly sensitive to finite size effects of the model slab in surface calculations.\cite{Hofmann2021} Surface energies and surface electronic states are often discussed as an example of this.\cite{Kratzer2019} Calculations of copper slabs with 30 layers or more are not unheard of.\cite{Hellsing2002}, as are corrections to address this slow convergence.\cite{Yoo2021}

% Fig: Sigma convergence  - all systems
\begin{figure}
    \centering
    \includegraphics{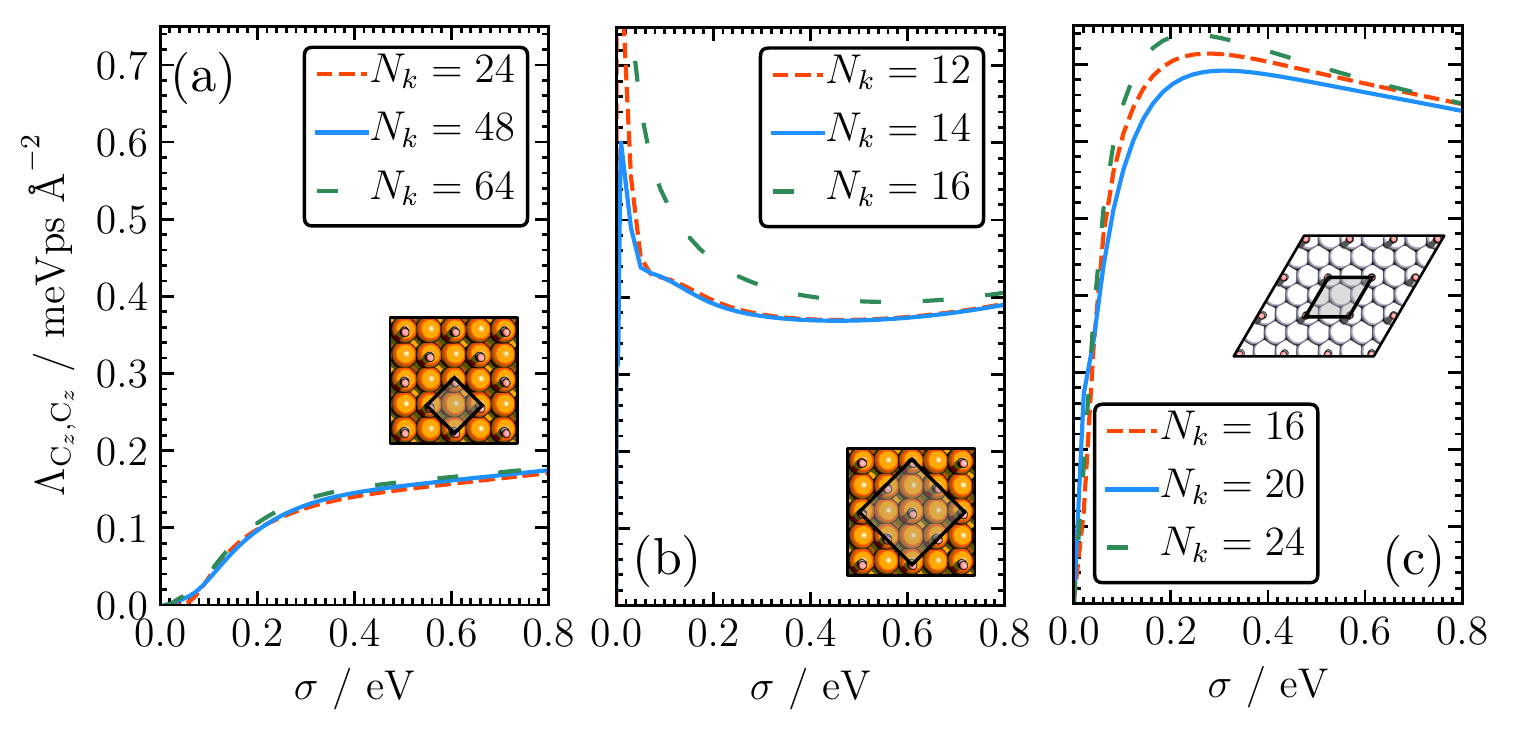}
    \caption{Behaviour of $\Lambda_{\mathrm{C}_z\mathrm{C}_z}$ with respect to delta function broadening $\sigma$ for the same systems as described in \cref{fig:tensor_layers_all}. The calculations here employ the delta function approximation given in \cref{eq:normalised_Gaussian} at zero frequency. Raw data used to produce this figure is available in the NOMAD repository under: URL to be added after acceptance.}
    \label{fig:sigma_convergence}
\end{figure}

\begin{figure}
    \centering
    \includegraphics{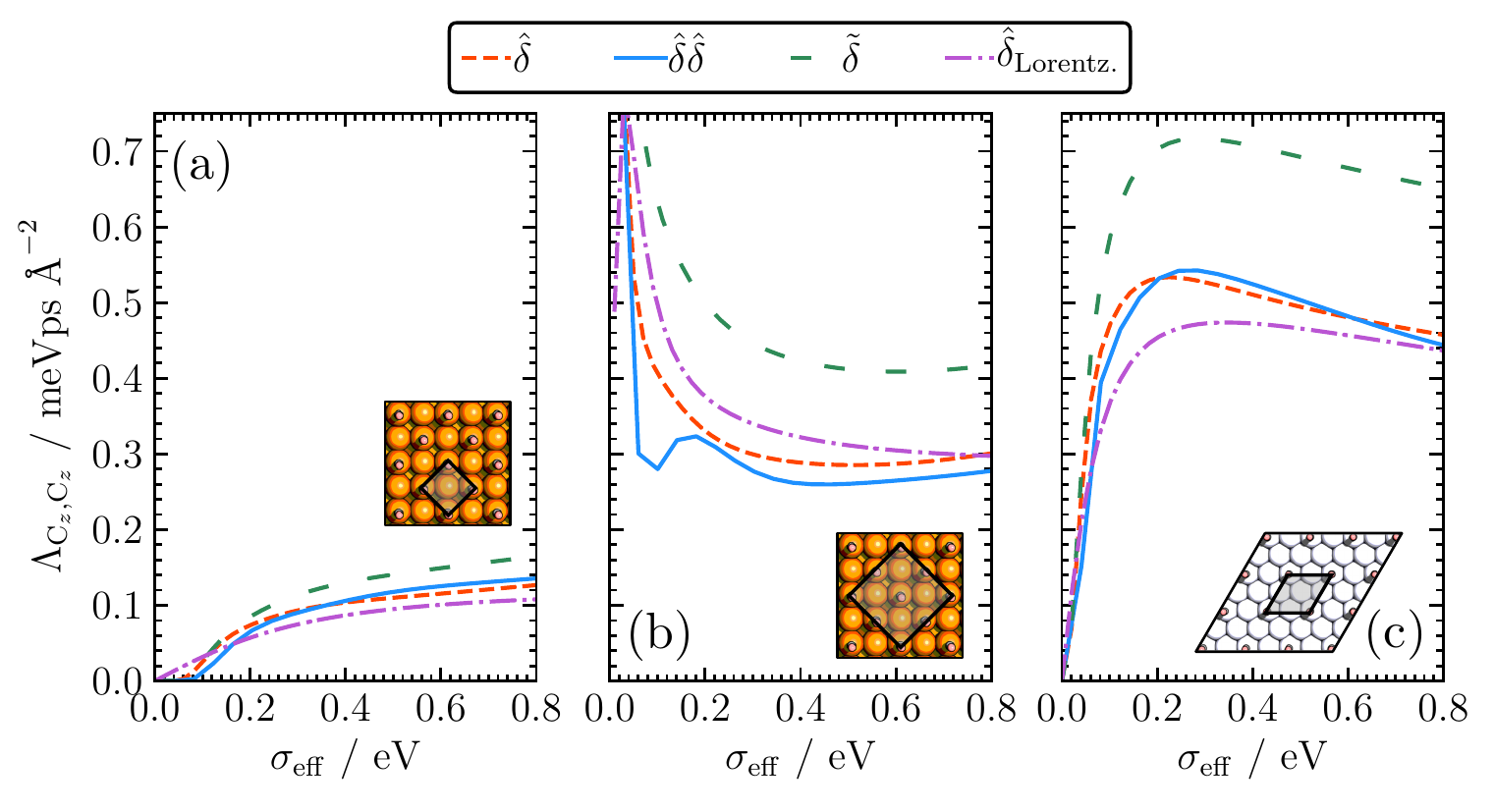}
    \caption{Comparison of single ($\delta$) and double ($\delta\delta$) delta function approximations which are defined above. Otherwise, same systems and settings as \cref{fig:sigma_convergence} except  with an electronic temperature of 0~K.  Zero frequency employed in the case of the single delta expressions. Raw data used to produce this figure is available in the NOMAD repository under: URL to be added after acceptance.}
    \label{fig:delta_types}
\end{figure}

% Par 6 - broadening dependence and delta type
We now explore the dependence of the friction tensor (via the $\Lambda_{\mathrm{C}_z\mathrm{C}_z}$ element) on the magnitude of broadening applied (\cref{fig:sigma_convergence}) and the type of delta function approximation (\cref{fig:delta_types}). All studied systems exhibit a plateauing behaviour once $\sigma>0.25$, however the profiles \cref{fig:sigma_convergence} differ in their behaviour for small broadening values between primitive cells and supercells (panel (a) and (b)). Small $\sigma$ values capture little to no friction for the primitive cells, even at sparser coverage p($2\times2$). This is expected as first-order EPC for $\q=0$ only capture interband excitations, which are not available in the zero frequency and zero broadening limit. \cite{Novko2016} The CO/Cu(100) $2\times2$ supercell captures significant friction with small broadening windows, though the dependence with respect to $\kk$-grid size appears to be less stable (panel b). The instability at small smearing widths is similarly reported by Shipley et al\cite{Shipley2020} (albeit for polyhydrides) when accounting for $\q>0$ excitations. The agreement between different $\kk$-grid sizes does improve at larger values of $\sigma$, however, excessively high values of $\sigma$ will incur errors and formally are not justified within first-order perturbation theory. \cite{Novko2016, Novko2018} 

Several delta function approximations are compared in \cref{fig:delta_types}. The single delta function approximation is roughly equivalent in magnitude for all smearing widths to the double delta function (considering that the effective smearing is double in the double delta function, since they are replaced by 2 Gaussians). The double delta function evaluation exhibits rather unpredictable behaviour at low smearing for the CO/Cu(100) supercell, whilst the single delta evaluation decays smoothly to a plateau. Despite previously being found to give more stable vibrational lifetimes for CO/Cu(100),\cite{Maurer2016a} application of additional normalization to the single delta function, described in \cref{eq:normalised_Gaussian}, appears to have no benefit in terms of stabilization with respect to $\sigma$ for any of the three systems, it does however significantly increase the friction tensor element. In terms of stability, the conventional single delta evaluation ($\hat{\delta}$) either via Gaussian or Lorentzian broadening appears to be preferential in terms of robustness and reproducibility. We note that the FHI-aims electronic friction module enables the use of all of the tested delta function modes, however, the single delta evaluation ($\hat{\delta}$, \cref{eq:friction_tensor_single_delta} with \cref{eq:Gaussian}) is the default option. All results where we compare to literature or experiment in sections and \ref{sec:benchmark} and \ref{sec:intraband} are generated with this option.

% Fig: Delta types
% Here we use sigma_eff ,  which is the number of delta functions * sigma. This is so we can compare single delta and double delta fairly.

%%%%%%%%%%%%%%%%%%%%%%%%%%%%%%%%%%%%%%%%%%%%%%%%%%%%%%%%%%%%%%%%%%%
\subsection{\label{sec:benchmark} Validation against pseudopotential plane wave DFT results}
% ---------------------------------------------------------------- %

In this section we investigate differences in electronic friction results obtained with the NAO all-electron representation in FHI-aims and the pseudopotential plane wave representation in {\sc Quantum ESPRESSO} (QE) codes, by reproducing two literature examples of molecules adsorbed on metal surfaces, reported by Spiering and Meyer \cite{spiering_testing_2018} and Diesen et al. \cite{diesen_ultrafast_2021}, namely H$_2$ on Cu(111) and CO on Ru(0001), respectively.
Spiering and Meyer performed electronic friction calculations with QE and EPW for the H$_2$ at Cu(111) surface to construct a neural-network-based electronic friction model to perform MDEF simulations of the vibrational state-to-state scattering probabilities. We also compare our H$_2$/Cu electronic friction values with the Luntz and Persson\cite{luntz_how_2005} (calculated with VASP), who investigated the impact of nonadiabatic effects on the activated adsorption/associative desorption processes, by studying two systems, H$_2$/Cu(111) and N$_2$/Ru(0001). Diesen et al calculated vibrational lifetimes for the internal stretch mode (IS), the surface-adsorbate mode (SA), and the frustrated rotation mode (FR) for the CO/Ru(0001), using QE code in order to validate experimental ultrafast dynamics results. 

% CO/Ru table
\begin{table}[]
    \centering
    \caption{Comparison of calculation parameters employed by Diesen et al\cite{diesen_ultrafast_2021} and this work for CO adsorbed on Ru(0001). Results are shown for both a 1 ML and 0.25 ML coverage. Both coverages were calculated with 6 substrate layers with the exception of values in parentheses which contain results calculated with 15 substrate layers. All calculations are performed at the $\Gamma$ $\q$ point only, with an electronic temperature of 0~K. Raw data used to produce this figure is available in the NOMAD repository under: URL to be added after acceptance.}
    %\small
    \footnotesize
    \begin{tabular}{lllllll} \hline
        Data Source & &$\gamma_{0,\mathrm{IS}}$  & $\gamma_{0,\mathrm{SA}}$ & $\gamma_{0,\mathrm{FR}}$  & $\Lambda_{dd}$  & $\Lambda_{ZZ}$ \\
        && / cm$^{-1}$ & / cm$^{-1}$ & / cm$^{-1}$ & / meVps\AA$^{-2}$ & / meVps\AA$^{-2}$\\
        \hline
        FHI-aims & 1ML & 0.59 (0.62) & 0.29 (0.40) & 1.19 (0.97) & 0.14 (0.14) & 0.10 (0.14)  \\
        QE & 1ML  &  1.15 (0.67) & 1.09 (0.67) & 1.29 (0.95) & 0.32 (0.19) & 0.45 (0.29) \\
         & 0.25ML & 2.01 & 0.85 &  1.28 & - & - \\        
        Diesen\cite{diesen_ultrafast_2021} & 1ML & 1.2 & 1.1 & 1.3 & - & - \\
          &0.25ML & 1.9 & 0.8 &  1.1  & - & - \\
         \hline
    \end{tabular}\\
    \label{tab:coru}
\end{table}

Table \ref{tab:coru} compares CO on Ru(0001) vibrational linewidths calculated in FHI-aims and QE with the results obtained by Diesen et al \cite{diesen_ultrafast_2021}. We calculate vibrational linewidths from the electronic friction tensor according to \cref{eq:broadening_with_friction} where the friction tensor is multiplied from left and right with the relevant mass-weighted normal mode displacement vector of the vibrational mode. The table also reports electronic friction tensor components projected along an internal coordinate related to internal stretch motion $\Lambda_{dd}$ and the molecular centre of mass motion $\Lambda_{ZZ}$. This internal coordinate representation is chosen to compare friction tensor values independently of normal mode calculations in both codes and is consistent with the one used below for H$_2$ on Cu(111) and previously used by Spiering and Meyer.\cite{spiering_testing_2018} When directly comparing the results of FHI-aims and QE with 6 substrate layers and the numerical settings as employed by Diesen et al., we find that there are significant discrepancies. The $\Lambda_{dd}$ component obtained in FHI-aims differs by a factor of 2, whereas the $\Lambda_{ZZ}$ obtained with QE differs by a factor of 4 from the FHI-aims value. This difference in major components of the friction tensor also translates into deviations in vibrational linewidths predicted by the two codes. Whereas the linewidth of the FR mode is described almost equally in both codes, both IS and SA components that feature contributions from Cartesian Z motion show larger linewidth broadening in QE than in FHI-aims when described with a 6-layer slab. 

% CO/Ru layer convergence with friction - figure:
\begin{figure}
    \centering
    \includegraphics[]{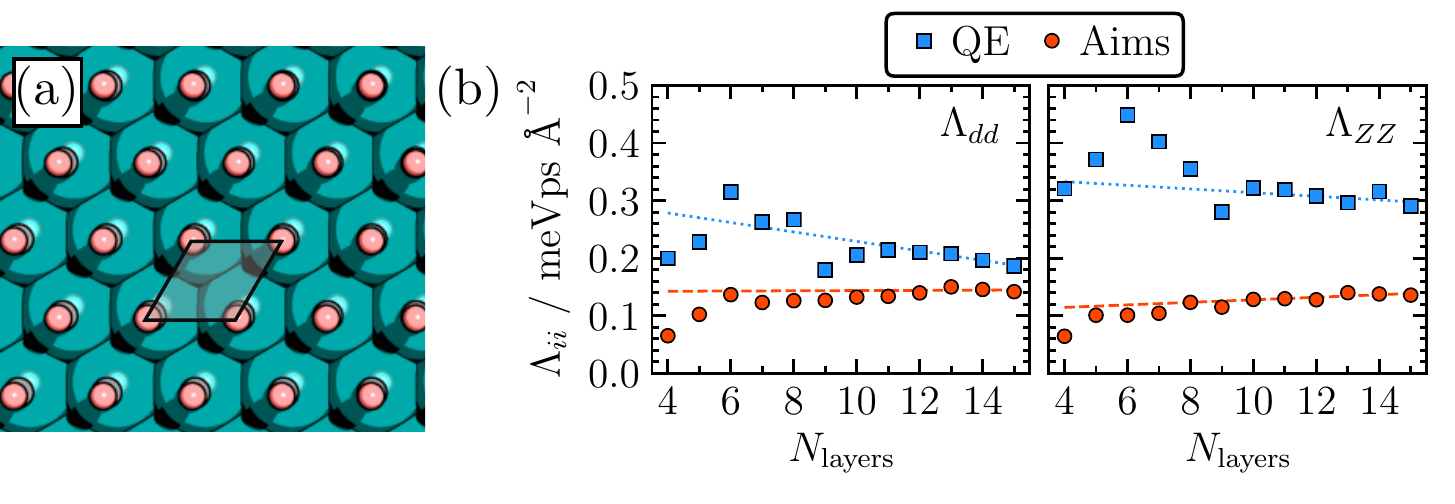}
    \caption{Graphical depiction of CO/Ru(0001) unit cell studied in this work (a) and layer convergence of 1 ML CO/Ru(0001) friction elements calculated with FHI-aims (Aims) and {\sc Quantum ESPRESSO}. Lines represent linear regression based on the last 4 points. Friction values are presented in internal coordinate representation. Raw data used to produce this figure is available in the NOMAD repository under: URL to be added after acceptance.}
    \label{fig:co_ru_nlayers}
\end{figure}

%CO/Ru convergence figure discussion and then back to table paranthesis values
We have already established in section \ref{sec:convergence} that several components of the friction tensor converge slowly with the number of substrate layers. The discrepancy between the two codes indeed arises from different convergence behaviour as can be seen in \cref{fig:co_ru_nlayers}. For both shown components of the friction tensor, FHI-aims shows monotonically increasing convergence with respect to layers, whereas QE shows large oscillations with a peak at 6 layers before approaching a converged value from above. Overall, the convergence is faster in FHI-aims with stable results already achieved at about 6-8 layers, whereas QE requires between 10 to 12 layers. Worryingly, even at 12 layers QE does not seem to be fully converged for both components. In the case of $dd$, both QE and FHI-aims appear to be converging to the same value. In the case of the $ZZ$ component, a significant discrepancy in values between QE and FHI-aims remains even for a slab with 15 substrate layers. While a further increase in layers may still reduce the gap between the values, we find it unlikely that the codes will converge to the same result. All settings in the calculations other than the number of substrate layers are fully converged with the only difference the type of basis representation and the use of pseudopotentials in the case of QE. We find that this discrepancy can be seen regardless of the type of pseudopotential that is employed. 
%We speculate that this might be related to a different response behaviour of the valence density perpendicular to the surface in the case of plane-wave pseudopotential calculations. 

Returning to Table \ref{tab:coru}, where the 15 layer results are shown in parentheses, we see that rigorous convergence eliminates all significant linewidth differences between the two codes except for the case of the $ZZ$ friction tensor component and the SA normal mode linewidth, which most strongly depends on the $ZZ$ component. It is important to note that the convergence settings used by Diesen et al. were more than sufficient within the context of their interpretations, but insufficient layer convergence of the electronic friction tensor may result in errors in the description of energy dissipation during MDEF simulations, which will become more apparent in the example of H$_2$ dissociation on Cu(111).

% H2/Cu friction along MEP - figure:
\begin{figure}
    \centering
    \includegraphics[]{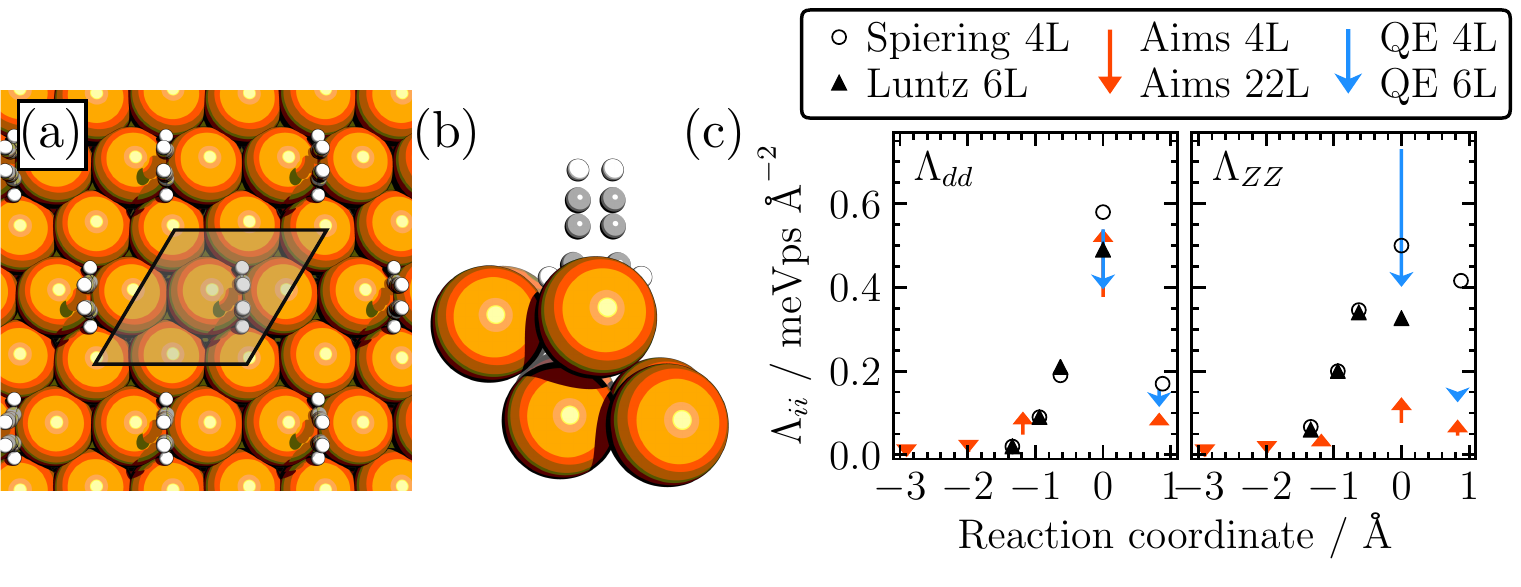}
    \caption{Graphical depiction of H$_2$/Cu(111) unit cell studied in this work (a), of the minimum energy path (MEP) of the dissociative adsorption of H$_2$ on Cu(111) (b), and electronic friction element values along MEP (c). Arrows depict the change in electronic friction values obtained with FHI-aims and {\sc Quantum ESPRESSO}, where arrow tails represent 4 layer systems and heads more converged systems containing more layers, ie. arrows point to better value reached with more layers. Circles represent previous results by Spiering and Meyer\cite{spiering_testing_2018} and Luntz and Persson\cite{luntz_how_2005}. The transition state is located at 0 \AA{}, the product state at 1 \AA{}. Friction values are presented in internal coordinates. Raw data used to produce this figure is available in the NOMAD repository under: URL to be added after acceptance.}
    \label{fig:h2cu_mep}
\end{figure}

Several previous works have reported electronic friction along the minimum energy path (MEP) of dissociative H$_2$ adsorption on Cu(111) as shown in \cref{fig:h2cu_mep}(a) and (b). \cite{luntz_how_2005,spiering_testing_2018} Following electronic friction along this MEP for the center of mass coordinate perpendicular to the surface, $ZZ$, and the intramolecular stretch, $dd$, shows that the friction tensor reaches its highest values at the transition state (TS), where the coupling with electron-hole pair excitations is the strongest. Our own calculations performed with QE and FHI-aims are able to reproduce this qualitative finding. \Cref{fig:h2cu_mep}(c) shows literature results together with our own calculations performed at comparable settings and with the highest number of layers we were able to include: 6 substrate layers in the case of QE and 22 substrate layers in the case of FHI-aims. We find that our 4 layer QE results at the transition state are in good agreement with the Spiering and Luntz results for $dd$, which were calculated with QE/EPW and VASP, respectively. Our results for $ZZ$ show a slight overestimation which may be explained by the use of different $\kk$-grid interpolation schemes, the Wannier approach by Spiering and the reciprocal space scheme implemented in PHonon module (QE) in our study. Our  FHI-aims results deviate significantly from the literature values and the QE results. The FHI-aims results are consistently lower in friction magnitude than what is found with plane wave calculations. In the case of the $dd$ component at the TS, FHI-aims with 4 layers slightly underreports the friction value, while the 22 layer calculation leads to an increased value that is inbetween what was found by Spiering and Luntz. Overall, increasing the number of layers in FHI-aims, does not affect the values of friction significantly, except in the case of the TS geometry. However, even with 22 substrate layers, the FHI-aims value for the $ZZ$ component at the TS geometry is less than a third of what Spiering and Meyer report and what our QE calculations provide. Similarly as in the case of CO on Ru(0001), by increasing the number of layers in the QE calculations, we see significant downward shifts in values, particularly for the TS geometry. Unfortunately, we were not able to generate QE calculations with a higher number of layers, but we expect that the gap between QE and FHI-aims will close further, as the Luntz result for $ZZ$ at TS calculated with 6 layers lies slightly below our own result at 6 layers.

% H2/Cu friction table
\begin{table}[]
    \centering
    \caption{Comparison of electronic friction values at the transition state ($\Lambda^{TS}$) and for the dissociated product state ($\Lambda^{Prod}$) of  H$_2$ dissociative adsorption on  Cu(111). Results obtained by Spiering and Meyer \cite{spiering_testing_2018}, Luntz and Persson\cite{luntz_how_2005}, and us with FHI-aims and {\sc Quantum ESPRESSO} (QE) are compared. Presented values refer to the p($2\times2$) unit cell containing 4 layers of Cu(111), except for the results reported by Luntz and Persson who used p(4x4) slabs with 6 layers of Cu(111) and values in parentheses which represent p($2\times2$) unit cell containing 22 layers of Cu(111). The unit of electronic friction values presented is meVps\AA$^{-2}$. Raw data used to produce this figure is available in the NOMAD repository under: URL to be added after acceptance.}
    %\small
    \footnotesize
    \begin{tabular}{lllll} \hline
        Data Source & $\Lambda_{dd}^{TS}$ & $\Lambda_{ZZ}^{TS}$ & $\Lambda_{dd}^{Prod}$ & $\Lambda_{ZZ}^{Prod}$\\
        \hline
        QE & 0.53 & 0.73 & 0.15 & 0.14 \\
        FHI-aims & 0.38 (0.53) & 0.08 (0.13) & 0.09 (0.09) & 0.05 (0.08) \\
        Spiering \cite{spiering_testing_2018} & 0.6 & 0.5 & 0.2 & 0.4\\
        Luntz \cite{luntz_how_2005} & 0.5 & 0.3 & 0.1 & 0.3\\
         \hline
    \end{tabular} \\
    \label{tab:h2cu}
\end{table}

Our findings on H$_2$/Cu(111) are summarized in Table \ref{tab:h2cu} and show that, once layer convergence is established, discrepancies between codes beyond what is expected from numerical differences are removed in the case of the $dd$ component (and other friction tensor components) with significant discrepancies only remaining for the centre of mass component, $ZZ$. The origin of such difference may lie in the use of a plane wave basis and the underlying pseudopotential (or projector augmented waves in the case of VASP) and will likely require further investigation. All the previously reported literature data for the systems studied in this section turned out to be underconverged with respect to the number of substrate layers when compared to the fully converged FHI-aims and QE results. Although we advise to investigate the layer convergence with any of the previously mentioned codes, we conclude that FHI-aims shows faster and more stable convergence behaviour for the calculation of electron-phonon linewidths and electronic friction tensors at metal surfaces.

\subsection{\label{sec:intraband} Interband and intraband contributions to vibrational relaxation of aperiodic adsorbate motion in c($2\times2$) CO overlayer at Cu(100)}
% ---------------------------------------------------------------- %

% Par 1 , intro to section, describe aperiodic vs periodic motion

In this chapter, we use the strong performance of the new code for a showcase application, namely to calculate the vibrational relaxation rate of a single CO molecule in a c($2\times2$) overlayer of CO adsorbed at Cu(100). The lifetime of the coherent ($\q=0$) CO overlayer on Cu(100) has been experimentally studied by IR absorption\cite{Ryberg1985a}, IR pump-probe\cite{Morin1992} and time-resolved vibrational sum-frequency generation spectroscopy\cite{Inoue2016}. In all cases, the coherent IS vibrational mode that is excited promptly decays via dissipation into electron-hole pair excitations and dephasing with other modes. There have been several attempts to theoretically reproduce this shortened lifetime based on calculating only the first-order electron-phonon lifetime of the $\q=0$ IS mode, \cite{Head-Gordon1992,Forsblom2007, Maurer2016a,Novko2016,Rittmeyer2017} all of which yield vibrational lifetimes and linewidths that are too small to be consistent with experiment or neglect other important contributions such as the coupling with substrate phonon modes. More recently, Novko et al. provided a comprehensive characterization of IS vibrational energy dissipation for this system, \cite{Novko2018,novkoUltrafastTransientDynamics2019} which included a description of second order electron mediated phonon-phonon coupling (EMPPC) that accounts for the coupling of the IS mode to other modes and the dephasing contribution to dissipation of the $\q=0$ IS mode. This description is fully periodic and includes how intraband excitations can lead to phonon-phonon coupling. The calculations we present in this paper do not include such contributions directly as we only consider first-order interband excitations, but intraband excitations can be included nominally by band folding when very large supercells are used, which our code enables.

% Graphical diagram showing aperiodic vs periodic q=0 vs periodic q=qmax
\begin{figure}
    \centering
    \includegraphics{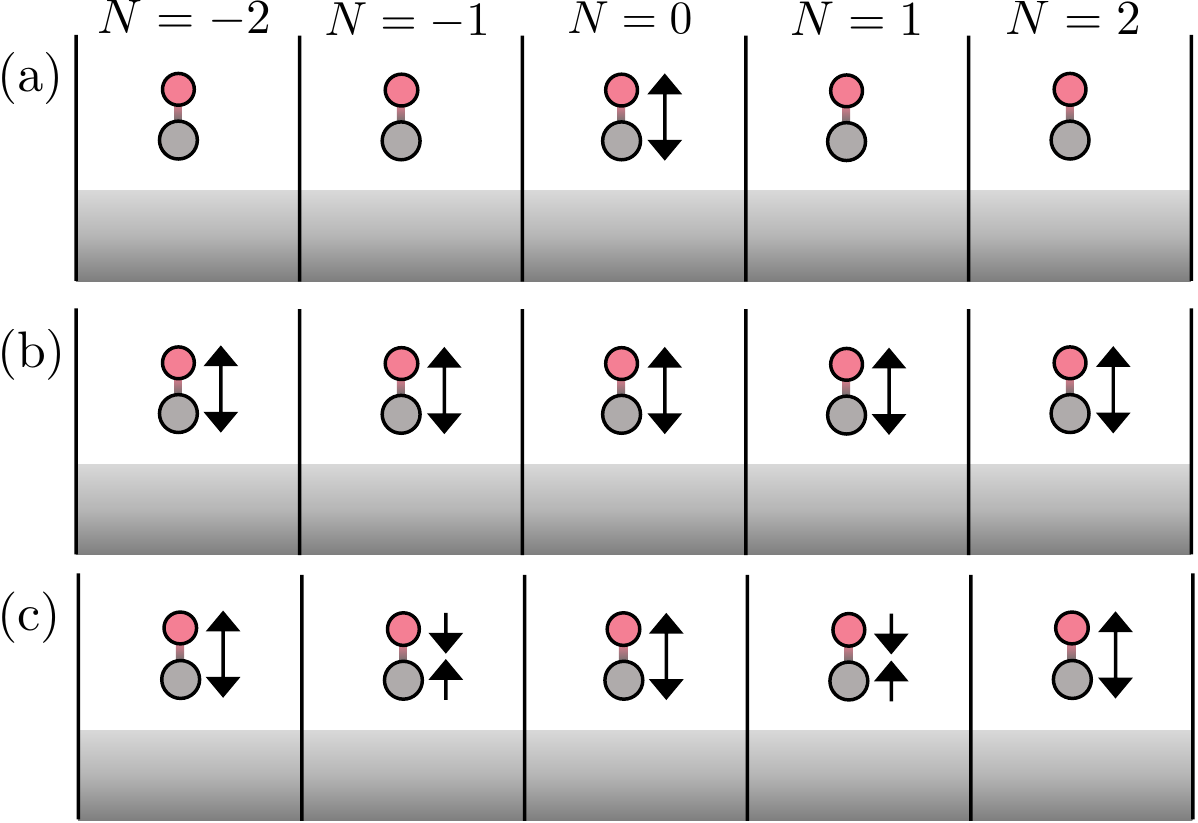}
    \caption{Graphical depiction of (a) aperiodic motion of adsorbate intramolecular stretch motion within a periodic surface overlayer, (b) periodic motion with $\q=0$ and (c) with $|\q|=q_{\mathrm{max}}$.  Crystal periodic motion is characterized by a momentum vector $\q$ and the associated perturbing potential can be represented by a phase modulated sum over displacement contributions of atom images in all periodically repeating unit cells. Aperiodic motion breaks translational invariance and can therefore only be approximated with an infinite Fourier sum over $\q$.}
    \label{fig:aperiodic}
\end{figure}

%  aperiodic motion albeit not directly comparable
Calculations presented in this work thus far have only considered electron-phonon coupling due to perfectly crystal-periodic motion. The corresponding perturbing potential is defined by a single crystal momentum vector $\q$ and the matrix elements and expressions follow as described in section \ref{sec:theory}. However, our aim is to be able to describe vibrational energy dissipation during surface dynamics in the context of heterogeneous catalysis, where typically individual adsorbates react at the surface and excite hot electrons. This is also the most relevant case for modelling molecular beam scattering experiments. We therefore differentiate between aperiodic and periodic motion on the surface in the context of periodic slab calculations. \Cref{fig:aperiodic} shows a graphical depiction of the aperiodic IS vibrational motion of a single adsorbate molecule in a periodic overlayer, the coherent totally symmetric motion of the $\q=0$ IS phonon, and the totally antisymmetric motion of the phonon with $|\q|=q_{\mathrm{max}}$.  The perturbing potential due to aperiodic motion (Fig. \ref{fig:aperiodic} right) of an atom in the principal unit cell can be described as a Fourier sum over all $\q$ vector contributions to the potential in the principal unit cell $N=0$:
\begin{equation}
    \frac{\partial}{\partial R_{a\kappa}}V = \sum_q e^{-i\q T(N=0)} \frac{\partial V}{\partial R_{\q,a\kappa}} = \sum_q \frac{\partial V}{\partial R_{\q,a\kappa}} 
\end{equation}

Let's consider coupling matrix elements that arise due to this aperiodic perturbation:
\begin{equation}\label{eq:EPC_aperiodic}
\tilde{g}_{mn, a\kappa}^{AP} (\kk) = \sum_q  \braket{m\kk+\q|\frac{\partial V}{\partial R_{\q,a\kappa}} |n\kk}
\end{equation}
Note that summation over $\q$ always implies normalization weighting with corresponding Brillouin weights $w_{\q}$. By moving the sum over $\q$ outside of the friction tensor description, we can see that the calculation of the vibrational relaxation rate for an aperiodic vibrational motion in the principal unit cell can be described as $\q$-averaging over the relaxation rates of all $\q$-modes in the irreducible Brillouin zone:
\begin{equation}
    \bar{\Gamma}_{\nu} = \frac{1}{N_{q}} \sum_{\q} \Gamma_{\nu,\q}
\end{equation}
As the code is currently restricted to $\q=0$ excitations, we perform this sum by increasing the size of supercells to include $\q>0$. This effectively folds intraband excitations into the principal Brillouin zone, which increases the number of available interband excitations. The improved memory distribution allows such systems to be calculated and the improved scaling of the code allows such cells to be tackled within the walltime of typical HPC systems.

%Now we expand our electronic states again into the AO basis:
%\begin{equation}
%        \tilde{g}^{AP}_{mn,a\kappa} (\kk) =  \sum_{\q} \sum_{ij}  \left(C^{j}_{m\kk+\q}\right)^* C^{i}_{n\kk}  \underbrace{\braket{\phi_{i\kk+\q} | \frac{\partial  }{\partial R_{\q, a\kappa}}V(\rr)|\phi_{j\kk}}}_{=h^{a\kappa}_{ij}(\kk,\q)}
%\end{equation}
%By considering the double Fourier sum, we arrive at:
%\begin{equation}
% \tilde{g}^{AP}_{mn,a\kappa} (\kk) =  \sum_{\q} \sum_{ij}  \left(C^{j}_{m\kk+\q}\right)^* C^{i}_{n\kk}   \sum_{N_e,N_p} e^{i(\kk\cdot\mathbf{T}(N_e)+\q\cdot\mathbf{T}(N_p))} \underbrace{\braket{\phi_{i}(\rr) | \frac{\partial  }{\partial R_{p a\kappa}}V(\rr)|\phi_{j}(\rr+\mathbf{T}(N_e))}}_{=h^{a\kappa}_{ij}(N_e,N_p)}   
%\end{equation}   

% Part 2 - Supercell convergence
% Intro to CO/Cu supercell and convergence 

% Figure CO/Cu(100) lifetime/relax rate/linewidth using supercells
\begin{figure}
    \centering
    \includegraphics[]{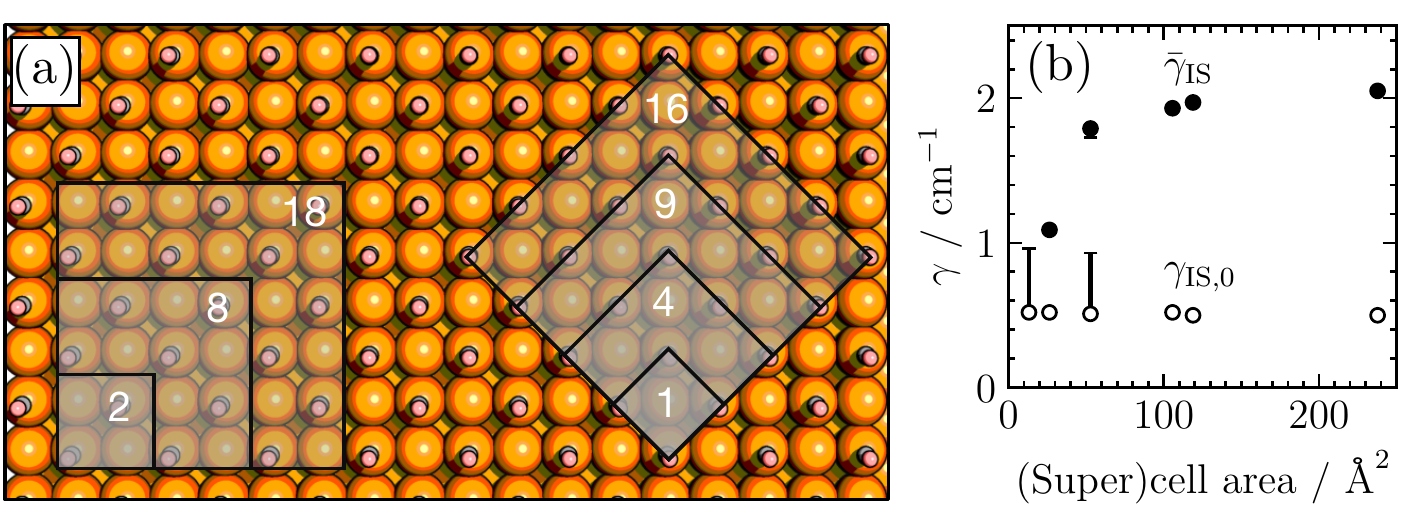}
    \caption{Graphical depiction of various supercells of c($2\times2$) CO/Cu(100) (a) and convergence of IS linewidth with supercell size (b) for $\q=0$ (open symbols) and $\q$-averaging (closed symbols). All points are calculated with 6 substrate layers, however the vertical bars show the recalculated values for a larger amount of layers, 16 for the primitive cell and 15 for the 3rd largest cell. The calculations here employ the delta function approximation given in \cref{eq:normalised_Gaussian} with $\sigma=0.6$~eV, evaluated at the zero frequency limit.  Raw data used to produce this figure will be available from a NOMAD link once published.}
    \label{fig:cu100_linewidth}
\end{figure}

In \cref{fig:cu100_linewidth}(a) we show a graphical depiction of various c(2x2) CO/Cu(100) supercells of different size, labelled by the number of CO molecules in the cell. The cells range from the primitive cell to a supercell containing 18 CO molecules, resulting in a $108\times 108$ friction tensor. For these cells, we calculate the $\q=0$ linewidth and the $\q$-averaged linewidth as shown in \cref{fig:cu100_linewidth}(b).  The symmetric IS linewidth ($\q=0$) is, as expected, largely independent of supercell size and matches that previously reported by Maurer et al\cite{Maurer2016a}. The $\q$-averaged linewidth increases with cell size until reaching convergence with supercell size at around 100 \AA{}$^{2}$ surface area. The increase is expected, as $\q>0$ modes contribute larger linewidths, \cite{Maurer2016a} and by increasing the unit cell more such modes are considered. The calculated normal mode frequency for the symmetric IS is 2059 cm$^{-1}$ for the largest cell, only 5 wavenumbers lower than the experimentally recorded frequency of 2064 cm$^{-1}$.\cite{Morin1992} The largest cell results are only possible with the improvements to the code and can act as a benchmark for future code development (i.e when including intraband excitations for smaller cells, which would be the preferred strategy when convergence is slow with respect to supercell size.)  We also show that the $\q=0$ linewidth is particularly sensitive to the number of substrate layers, by recalculating the linewidth for the primitive and a small supercell with 16 and 15 layers, respectively. The increase in linewidth with the additional substrate layers corresponds to roughly a 0.1 $\mathrm{ps}^{-1}$ difference in relaxation rate, which is consistent with the convergence plot in  \cref{fig:co_cu_supercell_convergence}. The q-averaged linewidth is stable with respect to number of layers for the supercell.

\begin{figure}
    \centering
    \includegraphics{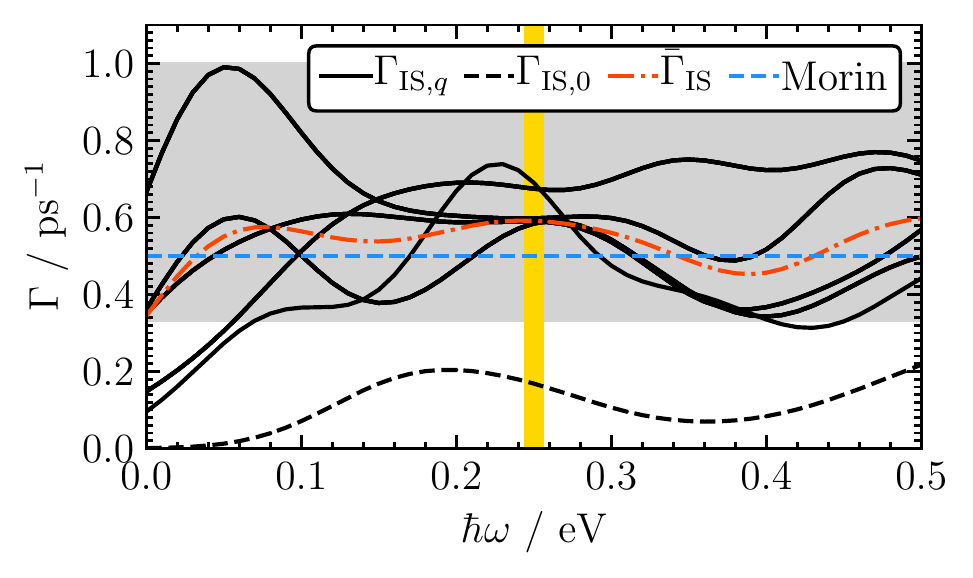}
    \caption{Energy dependent relaxation rate of the internal stretch mode for largest unit cell shown in \cref{fig:cu100_linewidth} for: $\q=0$ (dashed line), $\q>0$ (solid lines) and $\q$-averaged (dash-dot line). Horizontal dashed line represents experimentally recorded relaxation rate with the  experimental error represented by the grey shaded region.\cite{Morin1992}. Vertical yellow shaded region is the range of calculated normal mode frequencies for different $\q$-points of internal stretch. Delta function approximation given in \cref{eq:Gaussian} is used with $\sigma=0.05$~eV. Raw data used to produce this figure is available in the NOMAD repository under: URL to be added after acceptance}
    \label{fig:perturb}
\end{figure}

Within the converged supercell containing 18 CO molecules, we can now evaluate the relaxation rates of various periodic IS combination modes ($\Gamma_{\mathrm{IS},\q}$), and of the $\q$-averaged aperiodic motion, $\bar{\Gamma}_{\mathrm{IS}}$) of a single CO molecule within the overlayer. \Cref{fig:perturb} shows the relaxation rate for these cases as a function of perturbing energy. We find that the energy dependence of the relaxation rate for the aperiodic motion is weak in the limit of 0.05~eV Gaussian smearing and at perturbing energies above 0.05~eV, whereas individual modes show stronger dependence. The $\q=0$ mode consistently predicts a small relaxation rate that is inconsistent with experiment. When evaluating the relaxation rate at the perturbing frequency of the IS mode (0.245~eV) as per \cref{eq:friction_tensor_single_delta}, we find a relaxation rate of 0.18~ps$^{-1}$ for the $\q=0$ mode, corresponding to a lifetime of 5.6~ps, and 0.6~ps$^{-1}$ for the $\q$-averaged mode, corresponding to a lifetime of 1.7~ps. Whilst the aperiodic motion is not directly comparable to experiment in the case of the vibrational linewidth of c(2x2) CO on Cu(100), it can be seen as an instant dephasing approximation. This may be a good approximation, as Novko et al\cite{Novko2018} reported that $\q>0$ contributions dominate the IS linewidth. Our predicted lifetime of 1.7~ps is in good agreement with the 2~ps observed in experiment. \cite{Morin1992} As previously mentioned, the $\q=0$ relaxation rate will be roughly double what is reported in \cref{fig:perturb} with larger substrate thickness, but still less than half the experimentally recorded one. Our supercell results show the utility of a highly efficient implementation and can guide future development of the code, particularly when extending the code towards non-momentum conserving EPC matrix elements ($\q>0$). 

% Friction spectra has numerical normalisation but no extra Gaussian normalisation.
% 0.05 eV windowing used here

% ================================================================ %
                        \section{Conclusion}
% ================================================================ %

We have presented an all-electron numeric atomic orbital implementation of electronic friction and first-order electron-phonon response within FHI-aims with excellent scalability and memory efficiency that can compete with existing pseudopotential plane wave implementations of electron-phonon coupling. We employ the code for the calculation of vibrational linewidths and electronic friction tensor components of several molecular adsorbate overlayers on metal surfaces. The implementation is documented in detail and is featured in the most recent development version of FHI-aims (2021). We showcase the scalability and utility of the code by investigating the convergence properties of vibrational relaxation rates and linewidths due to electron-phonon response. We find that in many cases, in particular for CO adsorbed on Cu(100), substrate layer convergence with some Cartesian components of the electronic friction tensor is very slow, requiring up to 20 substrate layers when describing the system in a primitive unit cell.

By comparison with literature and calculations that we perform in the plane wave pseudopotential code {\sc Quantum ESPRESSO}, we find that our atomic orbital all-electron implementation provides a generally faster and monotonic layer convergence behaviour. The oscillatory convergence behaviour in plane wave calculations of metal-organic interfaces suggests that many calculations in literature on such systems may be insufficiently converged. Upon full layer convergence, we can show that plane wave pseudopotential and all-electron atomic orbital calculations agree for all contributions to the electronic friction tensor and linewidth, except for certain components perpendicular to the surface, where a discrepancy remains. We speculate that this discrepancy arises from a remaining difference in the response of the density in plane wave pseudopotential calculations, the origin of which will require further investigation. In our calculations, copper surfaces represent particularly difficult cases to achieve layer convergence for electron-phonon response, which likely is related to the low density-of-states of copper at the Fermi level. Finally, we showcase the capabilities of the code by describing aperiodic surface motion in large unit cells of unprecedented size by averaging over inter and intraband contributions for CO on Cu(100). We show that this ``instant-dephasing" limit yields a vibrational lifetime that is in good agreement with experiments\cite{Morin1992} and previous calculations in literature that consider higher order electron-phonon contributions.\cite{Novko2018}

The new implementation will enable the calculation of EPC-induced vibrational linewidths for complex systems with intrinsic length scales that require unit cells too large to be treated with existing plane-wave pseudopotential implementations. It will further enable on-the-fly ab initio MDEF simulations \cite{Maurer2017} and the large-scale generation of machine-learning training data to predict the electronic friction tensor as a function of atomic coordinates.\cite{Zhang2020} We have also shown an approach to calculate electronic friction and EPC linewidths for aperiodic motion within a periodic slab approach. In the future, we plan to extend this implementation to evaluate EPC matrix elements for $\q>0$ motion, which will likely require real-space interpolation approaches as previously proposed for EPC calculations with maximally localized Wannier functions \cite{giustinoElectronphononInteractionUsing2007} or atomic orbitals \cite{agapitoInitioElectronphononInteractions2018}. This will be particularly useful where the integration over $\q$-space converges too slowly to be addressed via supercell construction.

\section{Author Information}
\subsection{Corresponding Author}
Reinhard J. Maurer, r.maurer@warwick.ac.uk 
\subsection{ORCIDs}
Connor L. Box: 0000-0001-7575-7161 \\
Wojciech G. Stark: 0000-0001-6279-2638 \\
Reinhard J. Maurer: 0000-0002-3004-785X

\section{Supporting information}

\ack

This work received financial support through an EPSRC-funded PhD studentship, the ARCHER2 embedded Computational Software Engineering programme (ARCHER2 eCSE), a Leverhulme Trust Research Project Grant (RPG-2019-078), and the UKRI Future Leaders Fellowship programme (MR/S016023/1). 
We acknowledge computational resources provided by the EPSRC-funded HPC Midlands+ computing centre (EP/P020232/1) and the EPSRC-funded Materials Chemistry Consortium for the ARCHER2 UK National Supercomputing Service (EP/R029431/1). The authors acknowledge fruitful discussions with Christian Carbogno, Andrew Logsdail, Mariana Rossi, Volker Blum, Johannes Voss, Paul Spiering, and Elias Diesen.

%ackowledge helpful comminucations with Johannes Voss and Elias Diesen.

\printbibliography
%\bibliography{references}

\end{document}